%% file: main.tex
\documentclass[natbib=false,authordraft,review=false,timestamp=false,nonacm,screen]{acmart}

\AtBeginDocument{%
  \providecommand\BibTeX{{%
    \normalfont B\kern-0.5em{\scshape i\kern-0.25em b}\kern-0.8em\TeX}}}

\usepackage[datamodel=acmdatamodel,backend=biber,style=acmnumeric,defernumbers=true,sortcites]{biblatex}
\usepackage{tabularx}
\usepackage{pifont}

\usepackage{longtable} 

\include{edit-macros}

\usepackage{tabularx}
\usepackage{pifont} 
\usepackage{multirow}

\makeatletter
\let\blx@rerun@biber\relax
\makeatother

\begin{document}
\title{Sustainability in Computing Education: A Systematic Literature Review}


\author{Anne-Kathrin Peters}
\email{akpeters@kth.se}
\affiliation{
 \institution{KTH Royal Institute of Technology}
  \country{Sweden}
}

\author{Rafael Capilla}
\email{rafael.capilla@urjc.es}
\affiliation{
 \institution{Rey Juan Carlos University}
  \country{Spain}
}

\author{Vlad Constantin Coroam\u{a}}
\email{coroama@tu-berlin.de}
\affiliation{
 \institution{Technische Universität Berlin}
  \country{Germany}
}

\author{Rogardt Heldal}
\email{rogardt.heldal@hvl.no}
\affiliation{
 \institution{Western Norway University of Applied Sciences}
  \country{Norway}
}

\author{Patricia Lago}
\email{p.lago@vu.nl}
\affiliation{
 \institution{Vrije Universiteit Amsterdam}
  \country{The Netherlands}
}

\author{Ola Leifler}
\email{ola.leifler@liu.se}
\affiliation{
 \institution{Linköping University}
  \country{Sweden}
}

\author{Ana Moreira}
\email{amm@fct.unl.pt}
\affiliation{
 \institution{NOVA LINCS, School of Science and Technology, NOVA University Lisbon}
  \country{Portugal}
}

\author{João Paulo Fernandes}
\email{jpaulo@fe.up.pt}
\affiliation{
 \institution{LIACC \& DEI-FEUP, University of Porto}
  \country{Portugal}
}

\author{Birgit Penzenstadler}
\email{birgitp@chalmers.se}
\affiliation{
 \institution{Chalmers University of Technology}
  \country{Sweden}
}

\author{Jari Porras}
\email{jari.porras@lut.fi}
\affiliation{
 \institution{Lappeenranta-Lahti University of Technology LUT}
  \country{Finland}
}

\author{Colin C. Venters}
\email{C.Venters@hud.ac.uk}
\affiliation{
 \institution{University of Huddersfield}
  \country{U.K}
}

\renewcommand{\shortauthors}{Peters, et al.}

\begin{abstract}
Research shows that the global society as organized today, with our current technological and economic system, is impossible to sustain. We are living in the Anthropocene, an era in which human activities in highly industrialized countries are responsible for overshooting several planetary boundaries, with poorer communities contributing least to the problems but being impacted the most. At the same time, technical and economic gains fail to provide society at large with equal opportunities and improved quality of life. This paper describes approaches taken in computing education to address the issue of sustainability. It presents results of a systematic review of literature on sustainability in computing education. From a set of 572 publications extracted from six large digital libraries plus snowballing, we distilled and analyzed the 90 relevant primary studies. Using an inductive and deductive thematic analysis, we study 1) conceptions of sustainability, computing, and education, 2) implementations of sustainability in computing education, and 3) research on sustainability in computing education. We present a framework capturing learning objectives and outcomes as well as pedagogical methods for sustainability in computing education. These results can be mapped to existing standards and curricula in future work. We find that only a few of the articles engage with the challenges as calling for drastic systemic change, along with radically new understandings of computing and education. We suggest that future research should connect to the substantial body of critical theory such as feminist theory of science and technology. Existing research on sustainability in computing education may be considered as rather immature as the majority of articles are experience reports with limited empirical research.  
\end{abstract}





\maketitle

\input{1_Introduction}
\input{2_Background}

\input{3_Method}

\input{4_Results_RQ1}
\input{5_Results_RQ2}
\input{6_Results_RQ3}
\input{7_Discussion}
\input{8_Conclusion}

\printbibliography[title={Primary Studies},keyword=primaryStudy]

\nocite{*}
\printbibliography[title={Further References}, notkeyword=primaryStudy]

\newpage
\section*{Appendix}
\input{table-primary-studies}

\end{document}

%% file: edit-macros.tex
\usepackage[normalem]{ulem} 
\usepackage{xcolor}

\usepackage{ifthen}
\newboolean{showcomments}
\setboolean{showcomments}{true} 
\ifthenelse{\boolean{showcomments}}
  {\newcommand{\nb}[2]{
    \fcolorbox{gray}{yellow}{\bfseries\sffamily\scriptsize#1}
    {\sf\small$\blacktriangleright$\textit{#2}$\blacktriangleleft$}
   }
   
  }
  {\newcommand{\nb}[2]{}
   
  }


%% file: 1_Introduction.tex
\section{Introduction}\label{sec:1}
The overall integrity of the biosphere on Earth is under serious threat. There are several tipping points for the Earth’s climate system that we have already exceeded \cite{IPCC:2022}. The accelerating socio-economic trends of the past century~\cite{Steffen:2015aa} have put us on a path toward undermining four of the nine global planetary boundaries that define the basic conditions for human existence on Earth~\cite{steffen2015planetary}. According to the assessment of the Intergovernmental Panel on Climate Change (IPCC) from 2022, human-caused climate change has led to a range of strongly adverse effects. These effects include disproportionate warming of the arctic regions, acidified oceans that threaten coral reefs and those who depend on them worldwide, glacial retreat that threatens freshwater supplies for billions of people, sea-level rise that threatens coastal communities worldwide, and with extreme weather events that primarily affect poorer nations~\cite{IPCC:2022}. The report describes that many of the impacts of global warming are now already "irreversible", placing over 40\% of the world's population at high risk of the effects of climate change. Access to food, energy, and water is globally predicated on unsustainable practices, will be made even harder to sustain with the unfolding climate crisis, and must change fundamentally in the coming years to support the human population (c.f.~\cite{Hanjra:2010ud}). In parallel to these massive challenges to the biosphere that supports human life, social issues are just as pressing with more violent conflicts and democracy in global decline~\cite{House:2022uf}. 
%
%
Addressing our current unsustainable situation thus requires ``rapid, far-reaching and unprecedented changes in all aspects of society''~\cite{IPCC2018}. 

In this context, Information Technology (IT) is prominently but ambiguously positioned on these issues, with respect to, for instance, rapid decoupling of the economy from ecological footprints or strengthening democracies. Education is seen as having a great potential for change towards sustainability~\cite{UNESCO:2014}. Existing approaches to education for sustainability, however, are also limited in how they allow for change and provide agency to teachers and students to contribute to a better world~\cite{holfelder2019towards,facer2017noveltyHope}. 

With the present work, we explore the potential of information technology for sustainability as it is conveyed in education. Its aim has been to systematically review the existing work on sustainability in computing education, a field of research that has grown and developed into a research discipline in the past years. We seek to collect and synthesise existing work and provide threads for which to further advance the work on sustainability in computing education. Another goal has also been to bring together relevant work from different research communities engaging with the topics of sustainability, computing, and education and to consider various forums for publication. We analyzed existing research publications on sustainability and computing and education with the goal to understand:
\begin{enumerate}
    \item \textit{Conceptions} of sustainability, of the relationship between computing and sustainability, and of education for sustainability
    \item \textit{Implementations of education} including educational content, pedagogical approaches, and effects of education
    \item \textit{Research design} conducted into sustainability education, including its quality
\end{enumerate}

We develop our analysis as follows: Section~\ref{sec:2} provides various perspectives on the role of IT for sustainability and an overview of sustainability as an emerging field of research in computing. It also includes different perspectives on the role of education for change and previous research reviewing computing education for sustainability, explaining the gap in research that this work fills. Thereafter, in Section~\ref{sec:3}, we present the research questions for this systematic review along with the research method. Three sections follow, each of them presenting results to one research question, synthesizing takeaway lessons at the end of the section. In Section~\ref{sec:7}, we narrate the findings of our results, especially providing themes or questions for future research and education development, while also discussing the threats to validity of our study. Section~\ref{sec:8} concludes this systematic review.

%% file: 2_Background.tex
\section{Background: Sustainability, Computing, and Education}\label{sec:2}
In the following, we provide a brief overview of the evolution of research on sustainability in computing generally, and in computing education specifically. We refrain from providing nuanced conceptualisations as they are part of the results of our analysis.


\subsection{Sustainability and Computing}
The verb `to sustain' and the noun `sustainability' stem from the Latin `sustenere', which means both `to endure' and `to uphold' something~\cite{Venters2019}. Accordingly, `sustainability'  refers to the capacity of a\emph{system to endure} for a \emph{certain amount of time}~\cite{Venters2019}. 
Within the well-known current conceptualisation of sustainability, put forward by the Brundtland commission in 1987~\cite{Brundtland1987}, the system in question is Earth itself, which needs to be preserved for an unspecified period of time, but which extends many generations into the future. The Brundtland definition thus encompasses two aspects: distributive intragenerational justice (``the essential needs of the world’s poor, to which overriding priority must be given'')~\cite[37]{Brundtland1987}, but also intergenerational justice, for which the preservation of the biosphere is a prerequisite.

Given the scale and urgency of various environmental and societal issues, and the fact that computing technology protrudes ever more parts of our societies and economies, the idea to deploy computing technology to address questions of sustainability reaches relatively far back. Early efforts were concerned mainly with the environmental sustainability of computing itself (i.e., resource and energy efficiency), in a paradigm that has often been called ``green IT''~\cite{murugesan2008harnessing}. 
The broader, and arguably potentially more impactful paradigm, is to use computing to address energy and resource conservation in all the other domains of our societies; a paradigm known as ``green by IT''~\cite{hilty2012energy}. Digital transformations are assumed to free the economy from the shackles of the physical world and make economic growth fair and ethereal (c.f. \cite{Sachs:2019}). However, cryptocurrencies, for example, have such energy-devouring designs that they are far from ethereal~\cite{Digiconomist:2022vq}.
And while there are examples of successful and dematerialising digital substitutions such as partly virtualised conferences~\cite{coroama2012_twin-conference}, the substitution of e-goods for physical goods has not proven to yield an overall reduction of energy consumption~\cite{Court:2020aa}. 
In contrast, Court and Sorrell~\cite{Court:2020aa} question such purported savings: ``we cannot conclude that e-materialisation has delivered significant energy savings to date or is likely to do so in the future'' as they depend on the embedding of computing in social and narrative structures as well as on values that underpin attempts at sustainability in computing~\cite{Knowles:2014aa}.

Sustainability in computing is not limited to the environmental dimension. One example is the role of IT for strengthening or undermining democracy. Digital connectivity is seen as key to increased participation in 21st-century democracies through e-voting and citizen participation~\cite{Commission:ub}. 
However, the primary mechanisms for online connectivity are mediated through largely unregulated companies that promote toxic content and socially divisive business models that undermine the foundations for democracy~\cite{Zuboff:2015tj}. These fundamental tensions illustrate conflicting views about the nature and future of IT and indicate potential tensions about what it would mean for IT education to contribute to a societal transformation towards sustainable and just societies.  

The Karlskrona Manifesto for Sustainability Design~\cite{becker2012:karlskrona-short, becker2015:karlskrona-long} aimed to provide a focal point for establishing a common ground for the computing community to engage with sustainability by advocating a set of fundamental principles and commitments that underpin sustainability design. The principles stress the importance of recognising that sustainability is an explicit consideration, even if the primary focus of the system under design is not sustainability. 
Building on Penzenstadler~\cite{penzenstadler2014infusing}, it also advocates that sustainability should be viewed as a construct across five dimensions -- environmental, economic, individual, social, and technical -- and considers the potential long-term effects of systems.
Pham et al.~\cite{pham2020} proposed alternative dimensions including purpose, design aesthetics, integrative and legal aspects. 
With the exception of the legal aspects, it is an open question whether these are dimensions equivalent to the ones listed above.

Individual sustainability connects to education as it is about reaching personal sustainability goals including sustainability skills and competencies transforming the role of individuals in their current life. As stated in~\cite{mammas-and-pappas}: ``\emph{Sustainable individuals are characterized by creating harmony, interconnection, and relatively high levels of self-awareness in their values, thoughts, behaviors, and actions as well as cultivating continued individual growth in their physical, emotional, social, philosophical, and intellectual abilities.}'' 
For instance, developing individual sustainability with engineering students~\cite{barrella2018} aims at change as a combination of harmony, awareness, and behaviour to engage with sustainability principles~\cite{shields2002} and create a sustainable personality~\cite{mammas-and-pappas}.
Thereby, emotional and intellectual sustainability are key factors supporting the abilities of individuals across a wide variety of disciplines~\cite{UNESCO:2017aa}. 
%
While a focus on individual sustainability was until recently less common, this started to change by both analysing how the other sustainability dimensions can create a positive effect on individual and group well-being~\cite{penzenstadler2021take},
and also inversely how human factors impact, for example, software sustainability~\cite{imran2021}.

There are endless possibilities to engage with sustainability in computing and the field of research on sustainability and computing is growing and has matured. It covers a wide range of Green IT aspects but also IT supported sustainability initiatives~\cite{lago2015:cacm}. There have been methodological efforts to understand and assess the economy- and society-wide indirect effects of computing, which are broad, subtle, and can be both beneficial and detrimental to sustainability~\cite{coroama2020:assessing, bergmark2020:assessing}. Terms used to refer to the work, besides ``green IT'' and ``green by IT'', are ``Digital Sustainability'', ``Sustainable Computing'' or ``ICT for Sustainability'' (ICT4S)~\cite{hilty2015ict}.


\subsection{Education for Sustainability in Computing}
Reflecting the growth of computing and sustainability into an established field of research, there have been several efforts to also develop and describe education on sustainability in computing. Existing publications, however, are scattered across numerous disciplines and venues. Some work has been done from within the \emph{computing and sustainability} community. By contrast, \emph{computing education}, which is its own research field~\cite{simon2015}, paid so far relatively little attention to sustainability~\cite{pollock2019}. 
\emph{Engineering education} is another separate field of research, in which sustainability is being discussed, sometimes also including perspectives on computing.

Yet another field of research in which computing and computing education could possibly be discussed and that computing education could build on, is the field of 
\emph{education for sustainable development} (ESD),
which provides conceptions of education and its role for sustainability. Different analyses are made in this field as to what should change in education. The United Nations see the need for substantial changes to curricula and learning objectives, requiring a “profound transformation of how we think and act”~\cite[p. 7]{UNESCO:2017aa}, and a reorientation towards key sustainability competencies such as normative thinking, anticipatory thinking, and systems thinking in a wider sense (c.f. \cite{wiek2011key,UNESCO:2017aa}). Other work suggests making education less focused on training and detailed learning objectives and curricula~\cite{Osberg2020,holfelder2019towards}. An alternative role of education could be to become a democratic political space in which alternative futures are imagined~\cite{OsbergBiesta:2008,amsler2017contesting}. 

Work within education for sustainable development are critically discussed in a research field called ``critical ESD''~\cite{getzin2019shifting}. Recently, for example, Stein et al.~\cite{stein2022:end-of-world} propose a shift from ``education for sustainable development'' to ``education for the end of the world as we know it''. Another example is Holfelder~\cite{holfelder2019towards}, who criticises the focus on defining competencies in ESD as maintaining the status quo.
As expressed by~\cite{UNESCO:2017aa}, developing education for sustainability is thus far more complex than integrating sustainability into existing curricula.
The ways in which existing studies build on education research generally, and ESD specifically, is part of this review.   

Within engineering education, which only sometimes explicitly discusses computing education, calls have been made for doubling down on traditional engineering subject matter and problem-solving methods such as optimal resource use, Green IT, and educating future engineers on the minimization of harm from engineering (c.f.~\cite{murugesan2008harnessing,Mulder:2017aa}). At the same time, others have called for engineering education to be more challenge-based and trans-disciplinary (c.f.~\cite{Ashford:2004aa,Mulder:2012aa}). 

\subsection{Other Reviews on Education for Sustainability in Computing}

Only a few surveys have so far attempted to map the existing proposals of education for sustainability in computing, and their scopes have been rather limited. The authors in~\cite{Fisher2016} provide some examples of the integration of environmental and societal sustainability into computer science curricula at eight different universities, and~\cite{Boyle2004} describes a theoretical analysis of what sustainability education for engineering studies should consist of, given third-party literature. Neither, however, reviews the current state of practice in computing education in general. 
Three studies provide perspectives specific to national contexts: 
~\cite{Arefin2021} provides an Australian perspective and proposes guiding questions for program design, 
~\cite{Leifler2020} presents Swedish data on engineering degree program directors and their motivations for and challenges in integrating sustainability, and
~\cite{Sanchez2020} presents a Spanish survey and map of sustainability integration, including in computing. 
~\cite{Andara2018} presents a roadmap of activities for integrating sustainability in engineering education, based on a sustainability literacy survey, a literature review, and an expert panel on pathways for integrating sustainability. 
\cite{pollock2019} reviewed the current state of research addressing the issue of climate change in computing education by searching the ACM and IEEE databases, conducted expert interviews and collected data from the computing education community during a conference. 
Finally, \cite{thurer:2018wu} outlines a research agenda for studying sustainability integration in engineering in general, centering on four main themes: (a) implemented practice, (b) values and norms among teachers (often seen as “subjects”), and students (often seen as “objects”), (c) the roles of other stakeholders and (d) the assessments of outcomes or results. 

In our study, we address all four themes through the contributions 1-3 listed above in the Introduction.
In relation to the reviews above, our contribution is the first comprehensive study that addresses both the conceptions, implementations, and research designs that have been used in prior studies on sustainability integration in computing, with a broad definition of sustainability, learning objectives, and pedagogical methods that include values and norms, the roles of stakeholders and the assessment of outcomes.

%% file: 3_Method.tex
\section{Research Method}\label{sec:3}
\begin{figure}
    \centering
    \includegraphics[width=0.9\textwidth]{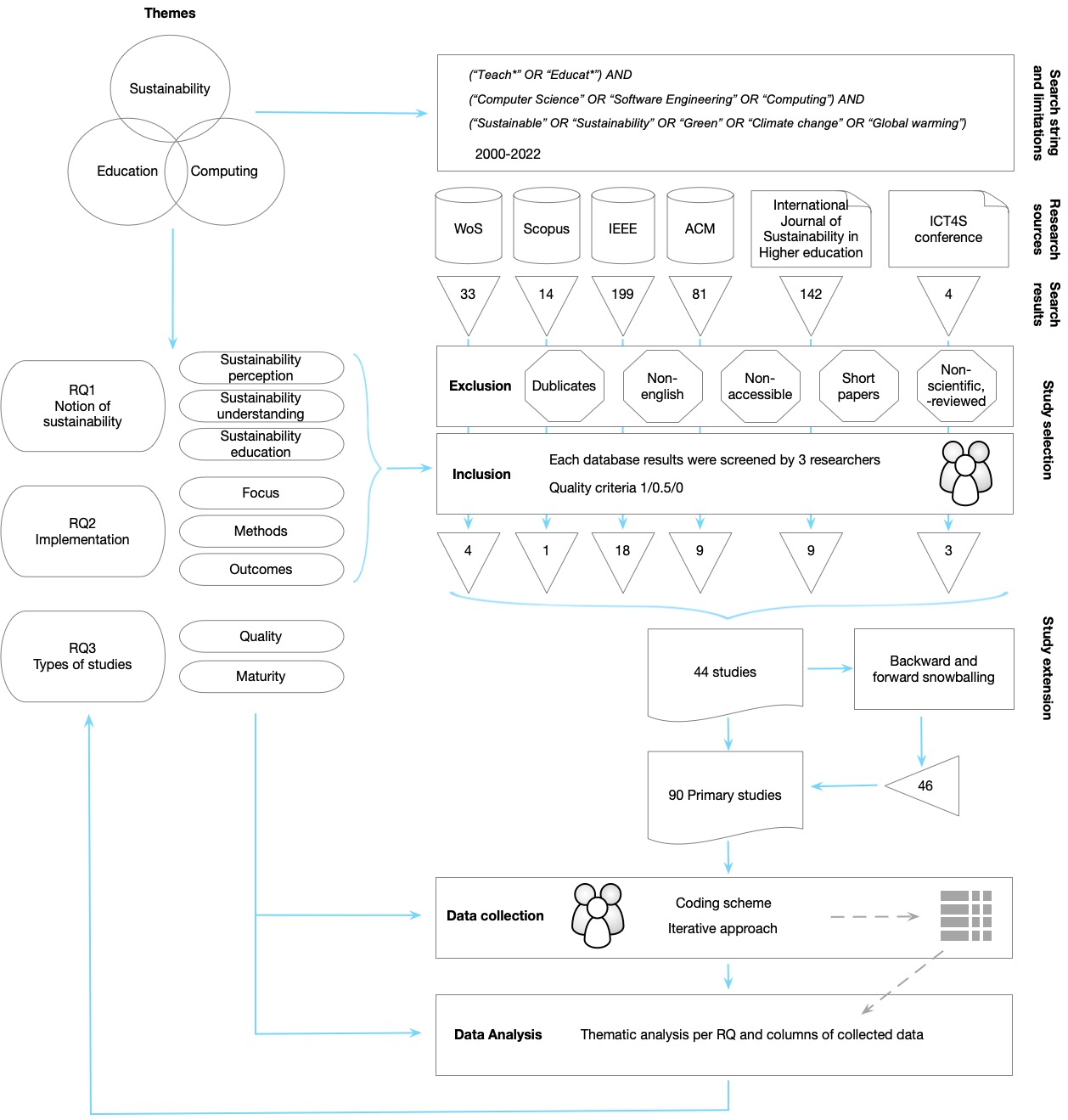}
    \caption{The literature review process followed in this study.}
    \label{fig:process}
\end{figure}

This section details the research questions, the methods we applied, and the process we followed. We use the term ‘computing’ as proposed by the ACM (2021) referring to an overarching discipline consisting of five sub-disciplines (ACM 2021): computer science, computer engineering, software engineering, information systems, and information technology. In this study, we follow the general guidelines for systematic literature reviews by Kitchenham \& Charters~\cite{kitchenham2007guidelines}, and the protocol for snowballing as presented by Wohlin~\cite{wohlin2014guidelines}. The general process is summarized in Figure~\ref{fig:process}. 


\subsection{Setting the Research Questions and Themes of the Study}
The following three research questions structured this work, each being specified through sub-research questions, as follows: 

\begin{itemize}
    \item [RQ 1] How does current research conceptualize sustainability? 
    \begin{itemize}
        \item [SRQ 1.1] How is sustainability being perceived? 
        \item [SRQ 1.2] How is the relationship between computing and sustainability understood?
        \item [SRQ 1.3] How is sustainability education being understood? 
    \end{itemize}
    \item [RQ 2] How is the implementation of sustainability education being presented?  
    \begin{itemize}
        \item [SRQ2.1] What content is sustainability education about, i.e. what are the learning objectives and the topics being taught? 
        \item [SRQ2.2] How is sustainability education being organized in terms of teaching practices and modules or curricula? 
        \item [SRQ2.3] What are the effects of teaching sustainability? 
    \end{itemize}
    \item [RQ 3] What type of research design has been conducted on sustainability and computing education?
    \begin{itemize}
        \item [SRQ3.1] What epistemological stance has been adopted as part of the research design?
        \item [SRQ3.2] What research methods have been adopted as part of the research design? 
        \item [SRQ3.3] What threats to validity are reported?
    \end{itemize}
\end{itemize}

Table~\ref{tab:themes-included-keywords} presents the themes and keywords that were identified from our research questions. They were used to create the following general search query. Different truncation (wild card) symbols were used in the searches to capture different word forms:  

\textit{(“Teach*” OR “Educat*”) AND (“Computer Science” OR “Software Engineering” OR “Computing”) AND (“Sustainable” OR “Sustainability” OR “Green” OR “Climate change” OR “Global warming”). }

\begin{table}[htb]
    \centering
    \begin{tabular}{l l}\toprule
       Theme & Proposed keywords  \\ \midrule 
       Education  & Education, Teaching, Learning \\ 
       Computing discipline  & Computer science, Software Engineering, Computing \\ 
       Sustainability & Sustainable, Sustainability, Green, Climate Change, Global warming \\ \bottomrule 
    \end{tabular}
    \caption{Themes included in our study with the selected keywords}
    \label{tab:themes-included-keywords}
\end{table}

\subsection{Defining Search Strategies and Selecting Research Sources}
In November 2020, we identified relevant papers by searching through the following databases, ACM, IEEE, Web of Science (WoS), and Scopus. We also considered the Emerald database, but due to constraints in the search tool, we only considered the Journal of Sustainability in Higher Education. We also searched the Springer database. However, with limited options to specify the search, the number of results was too large. In all cases, we searched the fields title, abstract, and author keywords for our specified keywords. To mitigate the limitations of the search, we did backward- and forward-snowballing~\cite{wohlin2014guidelines} to identify further relevant papers. We also decided to include relevant papers from the sustainability workshop of the ICT4S conference as the topics relevant to this study have been covered there. Table~\ref{tab:databases} lists the searched databases, the search string used, some limitations, the number of retrieved publications, and the number of selected primary studies for data extraction and analysis. We repeated the scan of literature in March 2022 to identify additional papers published since the first search. 

\begin{table}[htbp]
    \centering \footnotesize
    \begin{tabular}{l p{7cm} p{3cm} l l}\toprule
        Database & Search string & Other limitations & Results & Incl. \\\midrule 
        ACM & [[Abstract: teach*] OR [Abstract: educat*]] AND [[Abstract: "computer science"] OR [Abstract: "software engineering"] OR [Abstract: "computing"]] AND [[Abstract: "sustainable"] OR [Abstract: "sustainability"] OR [Abstract: "green"] OR [Abstract: "climate change"] OR [Abstract: "global warming"]] & AND [Publication Date: (01/01/2000 TO 27/3/2022)] & 81 & 9 \\
        IEEE & (("Abstract":teach* OR "Abstract":educat*) AND ("Abstract":"computer science" OR "Abstract":"software engineering" OR "Abstract":computing) AND ("Abstract":sustainable OR "Abstract":sustainability OR "Abstract":green OR "Abstract":"climate change" OR "Abstract":"global warming")) & Timespan: 2000-2022 & 199 & 19 \\
        WoS & TITLE: ((teach*  OR educat*)  AND (computer science  OR software engineering  OR 'computing')  AND (sustainable  OR sustainability  OR green  OR climate change  OR global warming)) & Timespan: 2000-2022 & 33 & 4 \\
        Scopus & TITLE-ABS-KEY-AUTH ( ( teach*  OR  educat* )  AND  ( 'computer  AND science'  OR  'software  AND engineering'  OR  'computing' )  AND  ( sustainable  OR  sustainability  OR  green  OR  'climate  AND change'  OR  'global  AND warming' ) ) & Timespan: 2000-2022 & 14 & 1 \\
        Emerald & ((teach* OR educat*) AND ("computer science" OR "software engineering" OR computing) AND (sustainable OR sustainability OR green OR "climate change" OR "global warming")) & International Journal of Sustainability in Higher Education & 142 & 10 \\
        ICT4S & Papers in the proceedings of the ICT4S conference series & International workshop on Computing+Sustainability+Education (CompSusTech) & 4 & 3 \\\bottomrule 
    \end{tabular}
    \caption{Digital databases, limitations, retrieved publications and selected primary studies} 
    \label{tab:databases}
\end{table}

\subsection{Selecting Studies Based on Inclusion and Exclusion Criteria}

The search results were collected in a spreadsheet, noting metadata such as title, authors, abstract, publication date. Study selection was done in two phases. First excluding papers by using the first five exclusion criteria and then including only papers fulfilling the inclusion criteria. Exclusion of the papers based on the exclusion criteria EC1-5 is simple as one does not have to focus on the contents. Only EC6 requires knowledge of the contents of the papers and as such, it was implemented as a part of the inclusion phase.

\begin{itemize}
    \item EC1: Duplicates from different data sources
\item EC2: Papers not written in English
\item EC3: Papers not accessible by the researchers
\item EC4: Short papers less than 4 pages long
\item EC5: Non-scientific and peer-reviewed publications, posters, thesis, or books
\item EC6: Papers from the same authors presenting the same results
\end{itemize}

In the inclusion phase the search results were categorized for relevance (quality criteria towards RQs). Two researchers rated the results independently, marking the results with either 1 (include), 0 (not included), or 0.5 (undecided). Undecided papers were further checked by a third researcher. Articles with a sum of at least 1 were included in our study. As a result, we included papers based on the following inclusion criteria.

\begin{itemize}
    \item 
IC1: Papers about computing and sustainability and education. We understand computing in a broad sense, including fields such as computer science, software engineering, and other related engineering disciplines addressing any sustainability dimension. (Papers that do not address education in computer science and other related engineering disciplines or don't include sustainability topics were not included.)
    \item
IC2: Papers in which sustainability is a goal for education and teaching. (If papers used sustainable teaching only to refer to the workloads, they were not included.)
\end{itemize}

The original search in the databases yielded 473 results. Reading the metadata (abstract, title, etc.), we found 46 articles relevant. Two of those articles were excluded after more detailed readings by several of the authors, resulting in 44 articles. 

\subsection{Study extension by snowballing}
\begin{figure}
    \centering
    \includegraphics[width=0.5\textwidth]{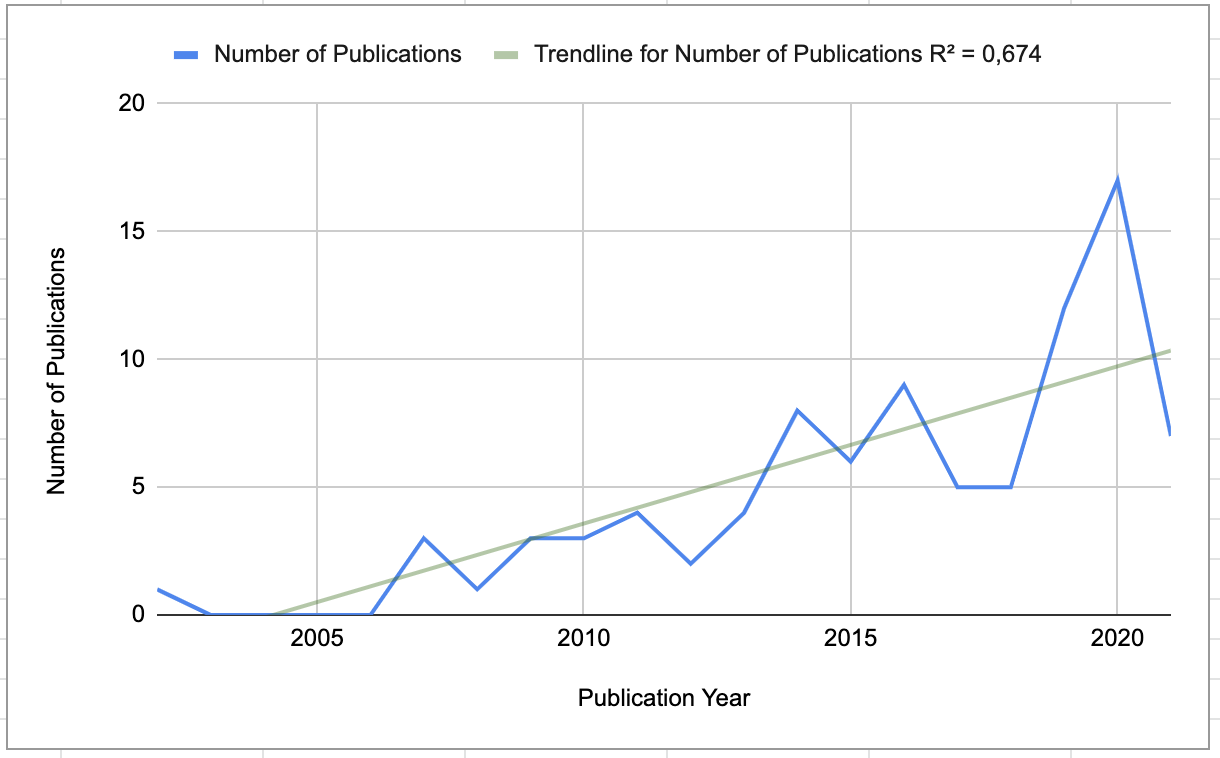}
    \caption{The selected publications per year.} 
    \label{fig:peryear}
\end{figure}

Backward and forward snowballing \cite{wohlin2014guidelines} with the selected 46 articles yielded another 99 articles that were identified as potentially relevant. Those articles were then checked for relevance by another researcher, resulting in 44 more relevant articles. Thus the final count of primary studies selected for our study was 90. Table \ref{long} in the appendix includes a list of primary studies, the type of publication venue and focus of the venue (education, computing, or sustainability). There is an increasing number of relevant papers over the years, with a majority of search results being less than 10 years old, see Fig.~\ref{fig:peryear}.

\subsection{Data collection}
To answer the research questions, we identified a set of themes to analyse the collected data, through a joint analysis of a small number of studies.  
We tested those themes, captured in a spreadsheet for the analysis of the paper, by analyzing the first papers and then discussing the use of the themes among the group. We then agreed on the following themes to analyse all the identified papers: 

\begin{itemize}
    \item Sustainability definition (perception of sustainability) [SRQ 1.1 and SRQ 1.2]
\item Understanding of what education of sustainability is [SRQ 1.3]
\item Learning objectives (goals) [SRQ2.1]
\item Best practices (means) [SRQ 2.2]
\item Implementation level (single course, minor, programme, curriculum, …) [SRQ 2.2]
\item Implementation techniques used in relation to recommendations for learning for sustainable development (see \cite{UNESCO:2017aa,2017Wals}).  Such techniques refer to learning methods reported to be helpful in developing strategic competencies such as normative thinking, systems thinking [SRQ 2.2]
\item Learning outcomes (students) [SRQ2.3] 
\item Study outcomes (long-term attitude changes, awareness change, feasibility of subject integration, faculty acceptance, faculty competence improvement) [SRQ2.3]
\item Study objectives [SRQ 3]
\item How and where were the objectives integrated into the curricula [Meta-analysis]
\end{itemize}

\subsection{Data Analysis} 
We performed a thematic analysis on the extracted results per theme and then consulted the referenced primary and secondary literature for more details and evidence as well as quotes. The data were analyzed in three smaller groups of authors that developed a coding scheme in several iterations of refinement. The findings were discussed in subgroups and synthesized into a narrative for reporting. During the whole analysis process, we have had weekly meetings to report and consolidate the results. 

To answer RQ 1, we developed the coding scheme through an inductive approach, identifying and revising themes capturing the different meanings in the data. We analyzed different ways of conceiving of sustainability, computing and sustainability, and education for sustainability in the papers. Oftentimes, we found such conceptions in the motivations for teaching and learning about sustainability in computing education. 

With regard to RQ2, we investigated how sustainability education has been implemented in CS and IT courses. We found the implementation described in terms of learning objectives, topics taught, pedagogical approaches and learning outcomes, for which we identified sub-themes in various iterations. 

For RQ3, we adopted a deductive approach to analyzing the kind of empirical research and the research quality in the primary studies In order to evaluate the quality of the studies according to the research method used, we adopted 
the ACM taxonomy of empirical standards~\cite{Ralph2020} to compare each primary study versus the different categories of empirical standards. As a result, we classified the different studies in any of the existing categories based on a number of essential attributes each type of paper must fulfill. Apart from this we also analyzed the epistemological stance of the approaches and if the the threats to validity found in the studies report incorrect conclusions. In this way we can identify the quality of the empirical studies reporting a particular case study or if they remain just only as proposals or vision papers.

In a further step, we synthesized the acquired results into an overall narrative and shed light on where the results are aligned and form a homogeneous picture and where it is heterogeneous and sometimes misaligned. We also identified and sorted out the overlap between the findings in the groups. Furthermore, we discussed the current state of practice in relation to what we know about best practices for education and education research for sustainability.

%% file: 4_Results_RQ1.tex
\section{Conceptions of sustainability, computing, and education (RQ1)}\label{sec:4}

In RQ1, the conceptions of sustainability and education were analyzed with respect to \textit{(i)} general conceptions of the current state of the world and what a sustainable development would mean, \textit{(ii)} how computing relates to the current state of the world and potential directions for a sustainable development, and finally \textit{(iii)} how education for sustainable development is conceived of in relation to the first two. Not all papers explicitly mention \textit{(i)} and \textit{(iii)}, but naturally all have conceptions of IT in relationship to sustainability. Conceptions are also part of some of the motivations in the papers, about why sustainability education in computing is important. For example, sustainability education in computing is described as important \cite{SB-16,SB-08}, as something that should be uncontroversial \cite{ACM-01}. Mann \cite{SB-05} argues that computing and education are two disciplines with high leverage for change with respect to sustainability. This implies a view of three disciplines that are to be connected for change.

We found several categories useful for distinguishing the conceptions described in the papers. However, they are neither intended to be exclusive nor comprehensive. A single paper may be included in several of the categories, so inclusion in one does not exclude from inclusion in others. Not all papers were explicit about their conceptions of sustainability, computing or education, so the list does not comprehensively capture conceptions in all papers. In some instances, a paper could feature several perspectives in one of the categories as well. However, these categories provide a lens through which we could understand the different conceptions and uncover the important differences. 

\subsection{Conceptions of Sustainability}\label{sec:4.1}
We identified six categories in our analysis of conceptions of sustainability. In some cases, those categories were explicitly referred to in the papers but in other cases, it was us as authors who interpreted the positioning of the papers from their identified descriptions in the papers.

The categories are:
\begin{enumerate}
    \item \textbf{Temporality}: present vs. future
    \item \textbf{Value transparency}: implicit vs. explicit
    \item \textbf{Originality}: common definitions vs. synthesis
    \item \textbf{Justification}: anthropocentric vs. ecocentric
    \item \textbf{Relativity}: weak vs. strong
    \item \textbf{Responsibility}: systemic and global vs. individual and local
\end{enumerate}

The \textbf{temporality} category makes a distinction between papers that primarily considered issues of sustainability-related to the \textbf{present}, and awareness of issues to address \cite{SB-89, SB-07, SB-33}, versus those that considered \textbf{future} states to be avoided (such as societal or ecosystem collapses) or desired (such as fulfilling the goals in Agenda 2030) \cite{SB-05}. For example, \cite{SB-89} noted that challenges of the present tend to be so-called wicked problems, not amenable to fixed solutions and optimal outcomes, but rather to be managed as social issues to be re-negotiated. This concerns the state of affairs in the present and issues we face. In such papers, awareness of the current state of affairs was considered important: “it is necessary that engineering students of all engineering disciplines be aware of environmental issues” \cite{SB-07}, or "\textit{green thinking} and the broad adoption of green software in computer science curriculum can greatly benefit our environment, society, and students" \cite{ACM-New-2}. Alternatively, some papers focus on actions for future directions that require radical changes to society: “Sustainability is considered here in terms of system change actions resulting in restorative socio-ecological transformation” \cite{SB-05}.

The \textbf{value transparency} category distinguishes between \textbf{implicit}, technical definitions of preserving the function of a system \cite{SB-01, ACM-06, ACM-12}, versus \textbf{explicit} value-based reasoning on world views, the rights or needs of people living now and those who live in the future, as well as other life on the planet \cite{SB-01, SB-31}. By technical definitions, we mean definitions that make value concerns invisible. They are also more technical in their choice of terminology. \cite{SB-01} provides a technical definition and describes that sustainability means to “use in a ternary relation as the use of a system S with regard to a function F and a time horizon L, which means using S in a way that does not compromise its ability to fulfill F for a period L” \cite{SB-01}. This follows from the tradition of System Dynamics and modeling the world in terms of systems of interacting components with feedback loops that can cause stabilizing, runaway or generally unpredictable behavior (see e.g., \cite{Ramage:2020aa}). However, as noted by systems thinkers such as Ulrich, values always inform the way we choose factual statements and system boundaries~\cite{ulrich1987critical}. The value-based definitions in the papers included \cite{SB-01, SB-31} mention of the New Ecological Paradigm (NEP) scale~\cite{Dunlap:aa} and the worldviews assumed to be associated with it: “we are particularly interested in understanding the deeper worldviews of the students ---a focus on the affective attributes of values, attitudes and beliefs” \cite{SB-31}.

Regarding \textbf{originality}, we saw differences between the re-iteration of \textbf{common definitions} in general literature on sustainability, and a \textbf{synthesis} of definitions produced by the authors themselves. Here, the definition from Our Common Future, also referred to as "Brundtland definition", appeared most often \cite{SB-01, ACM-12, ACM-11, SB-88, IEEE-163, IEEE-172, SB-71, ACM-14, SB-44, IEEE-New-2} planetary boundaries~\cite{Rockstrom:2009aa} appeared as a concept in two papers~\cite{IEEE-175,SB-45}, the Triple Bottom Line in one~\cite{Mann2011} (a term the original author has since tried to revoke~\cite{Elkington:2018aa}), Agenda 2030~\cite{:2015ab} appears in two papers \cite{IJSHE-26, SB-44}, and the three dimensions of economic, ecological and social concerns were mentioned in another three papers \cite{CompSusEd-02, SB-56, IEEE-New-2}. There were several examples of extensions regarding the relationship to technology, however. \cite{SB-40} provides a definition that specifically relates sustainability to the construction and use of technology: “Sustainability is an emergent phenomenon from the interaction between (very large numbers of) people, and the ways in which we build and use technology”. 
Others referred to the three dimensions of sustainability but also add individual and technical concerns \cite{IEEE-172, IEEE-New-1} \cite{penzenstadler2013generic}, or add a fourth ethical pillar \cite{IEEE-New-4}.

In distinguishing different \textbf{justifications} of sustainability, we could see both \textbf{anthropocentric} and \textbf{ecocentric} perspectives~\cite{Imran:2014aa}. From an anthropocentric perspective, sustainability is justified by the desires and needs of people. Environmental and social concerns are valid to the extent they concern the interest of human beings. An ecocentric perspective on the other hand justifies sustainability from the perspective of the inherent value of all life on the planet, treating life as sacred on its own terms. The nine papers that adopt the definition from Our Common Future~\cite{SB-01, ACM-12, ACM-11, SB-88, IEEE-163, IEEE-172, SB-71, ACM-14, SB-44} and the three that mention Agenda 2030 and the Sustainable Development Goals~\cite{IJSHE-26, SB-44, ACM-New-1} can be considered anthropocentric as Our Common Future centers on the needs of humans, 15 of the 17 SDGs concern humans, and only SDG 14 \& 15 list goals that pertain to all other life~\cite{Adelman:2018vi}. Two papers adopted an ecocentric perspective on sustainability \cite{SB-05, SB-31} as they describe sustainability as “restorative socio-ecological transformation” \cite{SB-05} and an “ecological worldview” \cite{SB-31}.

The \textbf{relativity} category contrasts \textbf{strong}~\cite{IEEE-175, SB-45, SB-07, SB-19, SB-62} versus \textbf{weak}~\cite{SB-01, IJSHE-13} (“vanilla”) sustainability \cite{neumayer2003weak}. The concept of strong sustainability is associated with respecting absolute, unconditional planetary boundaries and social foundations must be respected to avoid societal or ecosystem-wide collapse. The papers that mention the planetary boundaries framework and global limits to human activities were considered to adhere to the notion of strong sustainability. For instance, in \cite{SB-62} the authors write that “We have overstepped several planetary boundaries and risk overstepping several more \cite{steffen2015planetary}. We are about to reach limits as to the resources we can extract from the earth \cite{bardi2014extracted}, and the changes wreaked are by now so profound that they will likely last for a geological period of time”. Weak sustainability instead focuses on the relative balance between economic, social and environmental concerns. \cite{IJSHE-13} writes that “sustainability is about giving equal consideration to the economic, environmental and social aspects. There is growing consensus among researchers that conceptions of sustainability must include consideration of environmental, economic, and social factors” \cite{IJSHE-13}. An example of a study that explicitly compares these two perspectives is \cite{SB-19}, which states that vanilla sustainability concerns "picking low-hanging fruit in the form of energy efficiency, incremental technological innovations, or by applying human ingenuity [that] will allow us to continue to live as we do today" \cite{SB-19}. They argue that the strong form of sustainability, however, is required to face serious problems “such as climate change, planetary boundaries, future scarcity of nonrenewable resources (fossil energy, minerals), and the consequent challenges this scarcity will pose for our economic system.” \cite{SB-19}. “Strong sustainability” here represents a realignment of IT with a new societal project, one that assumes that fundamental changes of contemporary western lifestyles are needed to avert the worst of climate crisis, resource depletion, and ecosystem destruction.

Finally, the \textbf{responsibility} category concerns whether sustainability is considered something that must be understood and addressed as something \textbf{systemic and global} \cite{SB-01, SB-05} or \textbf{individual and local} 
\cite{ACM-06}. Here, there is not necessarily so much a conflict between different conceptions of what sustainability is, but rather what makes most sense to direct the attention of teachers and students to in the context of education. For example, \cite{SB-01} takes a systemic and global perspective and mentions that “The key challenges of the 21st century can mainly be characterized by their global impacts” \cite{SB-01}. \cite{ACM-06} on the other hand, begins by outlining global issues such as prolonged droughts that are the results of the climate crisis, but continue to focus more on individual action: “To become more resilient to future climate conditions \cite{adger2005successful}, create a greener society, save money, and improve food quality \cite{national2014garden}, many individuals are practicing home horticulture” \cite{ACM-06}.

Of the papers that make it apparent how they conceive of sustainability, we notice that the most prevalent conception is of sustainability as synonymous with Agenda 2030 and the definition in Our Common Future, which are anthropocentric and established in related literature. Ecocentric perspectives, in contrast, are much less common, although a number of different perspectives exist in individual papers.

\subsection{IT and Sustainability}
\label{sec:42}
During the analysis, we found three perspectives on the role of IT in sustainability: 
(i) an \textbf{incremental} perspective where the focus is on adjusting computing to reduce its footprint, (ii) an \textbf{enabling} perspective where IT is the means to address society-wide sustainability issues, and (iii) a \textbf{disruptive} perspective, which acknowledges the severe sustainability challenges we are facing that require ---as also argued e.g. by the IPCC\footnote{https://www.annualreviews.org/doi/abs/10.1146/annurev-environ-012220-011104}--- major, structural societal changes, and where also IT norms, practices and processes must change to align with and support the core principles of sustainability. 

Several primary studies focus on \textbf{incremental} changes within computing, adjusting the production of software and hardware in order to reduce their environmental footprint. Also beyond the education literature covered in our study, this paradigm has often been called “green IT” \cite{coroama2009energy, hilty2015ict} or “green computing”. This is in accordance with \cite{SB-01} that advocates the green use, disposal, design and manufacturing of IT systems as proposed by \cite{murugesan2008harnessing}. Several of the analyzed studies \cite{SB-01, ACM-11, ACM-12, IEEE-163, SB-77, SB-69, SB-45, SB-29, ACM-New-1, ACM-New-2} acknowledge the IT footprint and suggest adjustments to how we currently work. They acknowledge that energy and material consumption represent a problem and stress the impact that ICT infrastructures have on the environment due to their high energy demand (“3\% of global electricity consumption” \cite{SB-01}, greenhouse gasses  \cite{ACM-10, ACM-11}, raw materials required that also contribute to toxic waste \cite{ACM-10, ACM-11}, and e-waste produced \cite{IEEE-163}). Adjusting IT to contribute to sustainability is the direction of some other studies that suggest handling sustainability as yet another quality attribute, e.g. \cite{WoS-2, ACM-11}. 

We further found studies recognizing the value of computer science or IT as an \textbf{enabler} to address many of the society-wide sustainability issues. This paradigm, often addressed as “green by IT” \cite{hilty2015ict}, 
“I(C)T enabling effect”~\cite{malmodin2014life} or occasionally also as “ICT handprint”~\cite{WOS-5}, aims at using ICT to support sustainability throughout the entire society and economy; it applies computing to contribute to societal change towards sustainability; without, however, expecting radical changes of the society or economy themselves. Sometimes this paradigm of greening outside IT is also called “green computing”, e.g. in \cite{ACM-12}, although such terminology can easily lead to confusion with the previous category, “green IT”, which addresses greening within IT. The enabling paradigm widens the focus from the IT industry itself towards society and economy at large, promoting a “new sustainable model by mutually advancing economic, environmental, and social goals” \cite{ACM-12}. Similarly, \cite{CompSusEd-03} (p.2) notes that “we extend the focus of sustainability-related competencies and skills from the underlying digital technology to the needs of a digital society”. Also, \cite{IJSHE-26} recognizes the value of IT as an enabling technology, stating that “Computer science or IT, does not explicitly correspond to any of the SDGs but is nevertheless considered to be crucial for the fulfillment of each of them".  
Additionally, \cite{ACM-10} uses IT as an enabler when discussing how IT can “look outward to solve sustainability problems that IT isn't actually causing” (e.g., lights and room temperatures managed by IT systems, or transportation and logistics systems) and as a reporter when using data management and visualization to support those systems. 

The third case focuses on a \textbf{disruptive} perspective that puts sustainability first and entails the realignment of IT with a new societal project, \textit{demanding for radical changes} in our professional role through some or all of the following characteristics: responsibility, accountability, systems thinking, and affective reasoning, and perhaps activism (also supported by \cite{schendler2009getting}) and challenging political structures and norms. This is in accordance with the idea of transformation mindset by \cite{WOS-5} arguing that a transformation cannot be “met through marginal lifestyle changes~\cite{placet2005strategies}, instead we need urgent and ambitious changes \cite{thogersen2009simple}”. So, a change in professional behavior is in order. As \cite{IEEE-New-3} formulates it, IT professionals should have abilities to “understand the principles [of] complex systems, [...] reason about systems of values [...], and to deal with conflicts between the values of different stakeholders” while also having effective competencies for the “engagement, commitment and behavioral change towards sustainability”~\cite{IEEE-New-3}. In particular,  several papers emphasize responsibility, accountability, and the skills to contribute to solve complex problems with conflicting goals, e.g.~\cite{SB-34} stresses the need for creative solutions to complex problems and the importance of engineers being able to critically assess the implications of their professional actions. Such skills go beyond traditional computational thinking, demanding the ability to understand the circular nature of the world we live in. Systems thinking is also addressed by several authors (e.g.~\cite{SB-01, SB-40}) as the approach that “provides the necessary bridge from computational thinking to sustainability practice [...]”. It "provides a domain ontology for reasoning about sustainability, a conceptual basis for reasoning about transformational change, and a set of methods for critical thinking about the social and environmental impacts of technology"~\cite{SB-40}. Such disruptive perspective might mean challenging established computing norms, practices and processes and aligning them with the core principles of sustainability \cite{SB-07, SB-01, IEEE-New-3} (e.g., reduce, reuse, and recycle), “finding innovative ways to use ICT in business processes to deliver sustainability benefits across the enterprise and beyond”~\cite{SB-01}, and also considering the Karlskrona Manifesto\footnote{https://www.sustainabilitydesign.org}~\cite{IEEE-New-3}. 

\subsection{Education and Computing Education for Sustainability}\label{sec:4.3}
In the following, we describe how sustainability education in computing is understood and presented. We identify two categories for describing this understanding:
\begin{enumerate}
    \item \textbf{Context}: computing vs sustainability as the main context of learning
    \item \textbf{Purpose}: training vs emancipation
\end{enumerate}
The first category \textbf{context} reminds of the previous discussion of sustainability as something that is added to computing or as something that is treated as the context to understand computing within (see Section~\ref{sec:42}). Similarly, we find views of \textbf{computing education} as the main context of learning, having a right in its own, and sustainability education being “added” to computing education. We find many examples of incremental changes to education described in~\cite{ACM-01, ACM-12, IEEE-163, IEEE-164, IEEE-165, IEEE-167, IEEE-Penz2011CSEET, IJSHE-21, IJSHE-26, SB-16, SB-19, SB-30, SB-31, SB-45, SB-53, SB-54, SB-55, SB-62, 
WoS-1, WoS-2, SB-04}. For example, sustainability education can be included in lectures on software engineering~\cite{IEEE-Penz2011CSEET}, where e.g. sustainability can be discussed as a yet another quality attribute of a software system \cite{WoS-1}. Another example is promoting sustainability competencies \cite{wiek2011key, pacis2020key, jensen1997action}, which is argued for in \cite{SB-54, SB-31}. A full description of sustainability content, which is included in computing education is presented in Sections~\ref{sec:51} and \ref{sec:52}. On the other side of the spectrum, \textbf{computing education} as a whole is reimagined, realigned, or transformed to be about sustainability education, as explained by \cite{sterling2004higher}. That means that computing education is re-contextualised, sustainability is the core value of computing education~\cite{ACM-14}, it is at the “nexus” of computing education~\cite{SB-21}. Such a change of education to address sustainability is called “transformative”~\cite{ACM-12}. There are few examples of work with such an idea of sustainability education. An investigation from 2007 finds that only one engineering programme in the USA had been redesigned~\cite{SB-34}. Samuel Mann with colleagues develop computing education that centres around the learning objective to become a sustainable practitioner \cite{SB-05, WOS-5, SB-21, SB-31}. Porras et al.~\cite{IEEE-153, IEEE-172, IEEE-175} as well as Lago and de Boer~\cite{CompSusEd-03} present master programs that were developed from the ground centering around sustainability.

Transformative change is difficult to achieve in the current system~\cite{SB-45}. Adding courses or modules to the existing education is an easier and more straightforward approach to implementing sustainability~\cite{ACM-12}. Yet, inserting sustainability into a crowded computing curriculum can be experienced as a challenge \cite{ACM-12}, which means that the insertion of sustainability can become more of an injection, a “patch-on”~\cite{SB-45}, sustai~ability is “squeezed in”~\cite{SB-62}. It requires balancing different stakeholder interests~\cite{SB-30, IEEE-167}. However, even smaller modules such as a lecture, may lead to more transformative changes of computing education later on~\cite{IEEE-Penz2011CSEET}. 

The second category \textbf{purpose} is about what education aims to achieve. Education with a focus on \textbf{training} centers around equipping the students with certain skills, knowledge, or competencies, which also include attitudes or values. Holfelder calls this conception of education “training” and argues that education here is an instrument for attaining some specified objective~\cite[p.~944]{holfelder2019towards} 
Most papers engage with education as training. Section~\ref{sec:51} provides a detailed description of the learning objectives as they are specified in the papers. A few studies reflect about education as something different, which we call \textbf{emancipation}, a term also used in the literature on education for sustainability \cite{kaufmann2019building, getzin2019shifting}. Education here supports an inner journey and personal growth to forge the future with one's unique personal experiences and background. Samuel Mann argues for such an understanding of education \cite{SB-05}: “Rather than specifying a predetermined set of behaviors to describe sustainability within a discipline, instead we aim to take students on a journey of themselves identifying what it means for them to think and act as a sustainable practitioner”. Mann describes two student examples and suggests that technical content can be learnt “on-demand”, as needed by the student or the project. Such an idea of education would imply a change of power relation from lecturer as experts deciding on how to expose the student to sustainability and crisis, to lecturer and learner as more equal and co-learners \cite{SB-05, CompSusEd-03}. Lago and de Boer write about co-creation of real-world examples for local communities~\cite{CompSusEd-03}. Mann et al.~\cite{ACM-14} state a vision that graduates, practitioners and academics understand concepts of sustainability so that they can “evaluate, question and discuss their role in the world and to enable them to make changes where and when appropriate”. This vision emphasizes the individuals’ unique possibilities in the world, if they are seen and given a space in education. The Vrije University of Amsterdam, lifts three core values, that could be interpreted as emancipatory views of education~\cite{CompSusEd-03}: (1) students’ responsibility for their education, (2) the interdisciplinary and open character of the programme, and (3) students’ responsibility for issues such as “sustainability, ethics, privacy”.

\subsection{Takeaways for RQ1}
This section provides categories that help conceive of \textit{sustainability}, \textit{computing and sustainability}, as well as \textit{education for sustainability in computing education}. For each category, we contrasted different values such as ecocentric vs. anthropocentric, or computing first vs. sustainability first. Those values should not be seen as opposites or binaries that are clearly separated, but rather reflective of the various, partly overlapping, sometimes difficult to match, perspectives found in the papers. 

Given this plethora of conceptions of sustainability that differ along various dimensions, it is perhaps surprising that they ultimately boil down to only three main possible roles that IT could take in sustainability: an \textit{incremental} perspective focusing on reducing the direct environmental footprint of computing, an \textit{enabling} perspective in which IT is the means to addresses society-wide sustainability issues, and a \textit{disruptive} perspective, which acknowledges and responds to the need of unprecedented change.

Our review of conceptions of education suggests that most primary studies view sustainability as an important concern for computing professionals, but not in such a way that it would require radically new degree programs. A few authors, however, see that education needs to be reconceptualised to support change towards sustainable development, which we here presented in terms of emancipatory education. 

%% file: 5_Results_RQ2.tex
\section{Implementation of education for sustainability in computing (RQ2)}
\label{sec:5}

In RQ2, we investigate how education for sustainability in computing is being implemented. We do so by researching the following aspects: \textit{(i)} the intended learning objectives (i.e., what the students should know at the end of a course, see Section~\ref{sec:51}), \textit{(ii)} the topics, or contents through which the learning objectives are being realized (Section~\ref{sec:52}, i.e., the content of education), \textit{(iii)} the organization of education in terms of pedagogical approaches and educational structures (Section~\ref{sec:53}), and finally (iv) the reported effects (i.e., the achieved learning outcomes and other effects, Section ~\ref{sec:54}). We close with a summary of the main takeaways (Section~\ref{sec:55}).

\begin{figure}[htb]
  \centering
  \includegraphics[width=\linewidth]{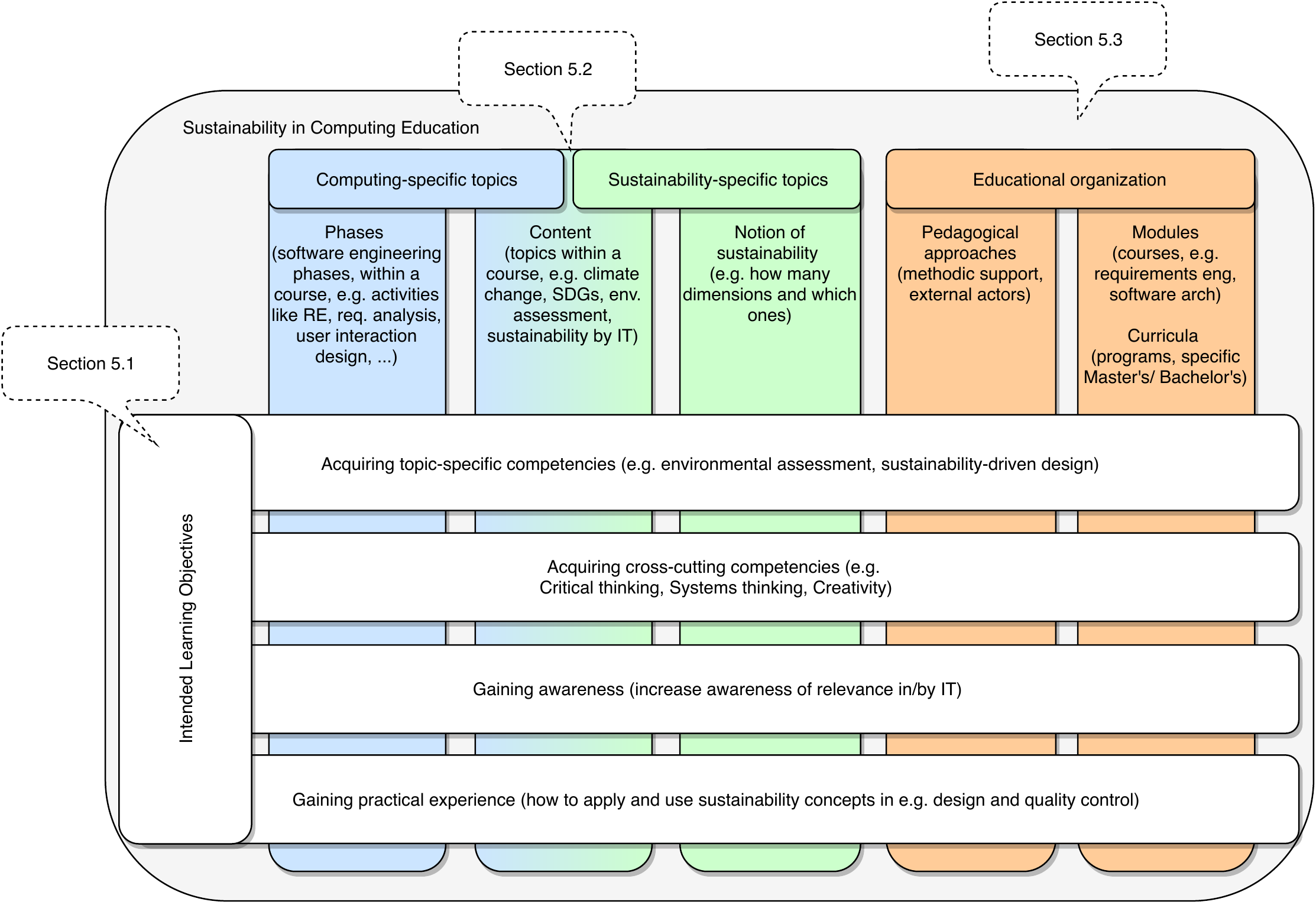}
  \caption{An overview of the different educational components in education for sustainability in computing}
  \label{fig:BigPicture}
\end{figure}

The systematic literature review results related to the aspects mentioned above are explained in the following sections and summarized in Figure~\ref{fig:BigPicture}. In this figure, the horizontal (white) pillars represent the types of intended learning objectives that were elicited from the primary studies. The topics are used to concretize learning objectives and are represented by the first three vertical pillars: they can be computing-specific (i.e., ICT topics where sustainability provides the application context), sustainability-specific (i.e., topics covering sustainability knowledge), or a combination thereof (i.e., topics that really mix both domains and as such cannot be classified as one or the other). Finally, the last two vertical pillars represent how education is being organized.

\subsection{Intended Learning Objectives (SRQ2.1)} \label{sec:51}

The intended learning objectives identified in the primary studies are what the students are expected to learn in a course or teaching intervention, i.e., the competencies the students should have acquired, or, in Bloom's terms, ``what students should know or be able to do at the end of the course that they couldn't do before’’ \cite{Bloom}. 
We organized the learning objectives in four broad categories: Acquiring topic-specific competencies, Acquiring cross-cutting competencies, Gaining awareness, and Gaining practical experience. In the following, we explain each category and map it on the corresponding Dublin descriptor(s)\footnote{The Dublin Descriptors are general statements about the ordinary outcomes that are achieved by students after completing a curriculum of studies and obtaining a qualification. They are neither meant to be prescriptive rules, nor do they represent benchmarks or minimal requirements, since they are not comprehensive. The descriptors are conceived to describe the overall nature of the qualification. Furthermore, they are not to be considered disciplines and they are not limited to specific academic or professional areas. The Dublin Descriptors consist of the following elements: Knowledge and understanding; Applying knowledge and understanding; Making judgments; Communication skills; Learning skills.  The European Higher Education Area \url{http://ehea.info} ~\cite{Gudeva2012-mo}}. This way, given the widespread adoption of Dublin descriptors in educational frameworks especially, but not exclusively, in the European Union, we aim to ease their use.

\textbf{Acquiring topic-specific competencies.} Topic-specific competencies are specific to the domain of computing and sustainability. In any educational component (module or curriculum), the intended learning objectives and the related topics are intertwined (as the second is instrumental to achieve the first). Therefore, the topics elicited from the studies are presented in Section~\ref{sec:52}. 
Depending on the types of competencies covered by the study, acquiring topic-specific competencies can be generally mapped on Dublin descriptors ``Knowledge and understanding’’, and ``Applying knowledge and understanding’.

\textbf{Acquiring cross-cutting competencies.} These competencies are relevant across applications and domains and allow people to act effectively in the target application or domain. Previous work has formulated such competencies in terms of ``key sustainability competencies'' which include, e.g., systems thinking and integrative problem solving \cite{wiek2011key, pacis2020key}. We see reference to the work by Wiek et al. \cite{wiek2011key} in \cite{SB-54,IJSHE-8,SB-12,SB-54,SB-71,WoS-1,SB-99}. Mann et al. \cite{SB-31} link to \textit{action competence\footnote{I.e., an individual’s capacity of critically selecting and conducting possible actions that may solve societal problems through democratic mechanisms (\url{https://wikieducator.org/Learning_and_Teaching_in_Practice/Action_competence}).}} in environmental education \cite{jensen1997action}.
In the primary studies, we identified the following list of cross-cutting competencies:
\begin{itemize}
\item \textbf{Systems thinking.} Systems thinking is a holistic approach to thinking in systems and included in various ways and contexts: for climate change and gamification \cite{ACM-01, CompSusEd-01}, in general to address sustainability in broad contexts \cite{CompSusEd-02,SB-01}; for problem solving in a systemic way \cite{IEEE-161, IEEE-165}, to include environmental-friendliness in computing processes \cite{IEEE-167}; and for balancing ICT, economics and sustainability \cite{IEEE-171, IJSHE-10, CompSusEd-02}. 
Also, systems thinking is mentioned as an instrument to acquire a reflective, holistic way of thinking \cite{IJSHE-6}. Easterbrook \cite{SB-40} proposes it in the form of a Toolkit with a Domain Ontology for Sustainability Thinking, Theories of Change, and Tools for Critical Analysis. Hilty and Huber \cite{IJSHE-11} conclude that ``teaching systems thinking (\dots) plays an important role in teaching sustainability, in particular when the fields of ICT and Sustainable Development (SD) are to be combined”. Specifically the ``rebound effect”, which can be used as a potential introductory example to systems thinking, provoked students to think about technical efficiency in a new way. 

\item \textbf{Multi-perspective thinking.} Multi-perspective thinking uses systems thinking to especially consider the concerns from various parties. For example, Krogstie and Krogstie \cite{SB-54} promote it as way to address the concerns of individuals versus those of social communities, e.g., by using personas and associated scenarios to explore different needs.

\item \textbf{Transformation mindset.} Acquiring a transformation mindset is proposed as ``a way of thinking that leads to transformational acts resulting in socio-ecological restoration’’ \cite{WOS-5}.

\item \textbf{Judgments.} Making judgments is another frequently-mentioned cross-cutting skill for many sustainability-related purposes: we make judgments in ethics \cite{CompSusEd-03}; we make judgments on the societal role of informatics in shaping the future, and in doing so we assume an active role as software professionals \cite{IEEE-New-3}; we use judgments to reflect on \textit{(i)} ethical and sustainability aspects of software systems, and \textit{(ii)} choices in terms of both product impacts and processes \cite{IJSHE-12}. We also use judgments for critical thinking (e.g., for exploring the use of technology to protect the environment \cite{IJSHE-5}; and for learning and reflectivity \cite{SB-99}).

\item \textbf{Creativity.} Creativity is reported as a key competence in software engineering for sustainability \cite{IEEE-161}; for which it found as especially necessary ``[\dots] for developing new insight, making new connections, and identifying new solutions''.

\item \textbf{Communication.} Sustainability in given subjects requires communication skills, included in primary studies \cite{IJSHE-12,IJSHE-5} as a skill to report orally and/or in writing. One paper uses specifically academic writing to communicate basic tenets of sustainability and IT in society, responsibilities of IT professionals, and the effects that a particular IT system has from a systems perspective \cite{IEEE-New-3}.

\item \textbf{Ecological approach.} The ability of applying an ecological approach to computing is proposed by some studies as a learning objective instrumental, e.g., to the creation, research and exploitation of ICT \cite{IEEE-167}, to create eco-friendly technology and processes \cite{IEEE-164}, and in general to develop STEM reasoning skills in socio-technical areas \cite{IJSHE-10}. A special angle is addressed by Qureshi \cite{IJSHE-8}, to develop the knowledge of how to develop and improve sustainable living (where ICT is a supporting instrument).

\item \textbf{Meta-learning.} ``Learning to learn’’ is a skill to ``support students’ development in a range of analytical, communication and learning skills appropriate to the subject matter and level of the course’’ \cite{IJSHE-5}. 

\item \textbf{Research methods.} Research methods that specifically benefit ICT for Development (ICT4D) and the SDGs are taught to the students. For example, Lago and de Boer \cite{CompSusEd-03} propose Green Labs and ICT4D Field Work in their preview of a Master program geared around Digital Society and Sustainability. Müller et al. \cite{IJSHE-7} propose a model combining education on sustainability with thorough training in scientific research methods, including example projects from an existing social science curriculum and its integration into a real-world laboratory on sustainable energy use.

\item \textbf{Critical thinking}. The concept of critical thinking goes back to the American philosopher John Dewey (1910), who defined it as ``active, persistent and careful consideration of any belief or supposed form of knowledge in the light of the grounds that support it, and the further conclusions to which it tends” \cite{Dewey1910}. Critical thinking methods are important for all students such that “the next generation of professionals may be more likely to give consideration to issues (e.g., the environment, social equity, biodiversity) that they otherwise would have ignored" \cite{SB-34}. 
Critical thinking methods have gained popularity \cite{CompSusEd-03, SB-62, SB-75}.
Another example of critical thinking is proposed by \cite{IEEE-165}, which uses Felber’s model for the common good matrix to help elicit values in an engineering project.
\end{itemize}

Depending on the types of competencies covered by the study, Acquiring cross-cutting competencies can be generally mapped on Dublin descriptors ``Make judgment’’, ``Learning skills’’ and ``Communication’’.

\textbf{Gaining awareness}. The need to create awareness, in the students and various stakeholders, of the role of software (and ICT in general) for sustainability is common to many studies. Awareness is often mentioned with a broad connotation, as stimulating the reflection on the role of software \cite{ACM-10, ACM-14, ACM-07, SB-44}; the role of informatics in society, legal issues and ethical professional responsibility \cite{SB-82}; and the role of ICT industry and its societal and environmental implications \cite{IEEE-171}; or the conceptual understanding and awareness of environmental issues \cite{SB-07}.
Lopez et al. \cite{IEEE-165} propose a framework to help the students think about the sustainability of their own thesis project. 

Further, many studies aim at creating awareness of the specific types of effects. For example, the mentioned enabling effects (i.e., SE for sustainability) include the impact of ICT on society \cite{CompSusEd-03}, and case-study specific ICT impact on sustainability \cite{IEEE-Penz2013cseet}. Studies that explicitly target both direct (e.g., software energy footprint) and enabling effects include, e.g., user participation and inclusivity \cite{CompSusEd-02}, creation of ICT solutions in resource-scarce contexts \cite{CompSusEd-03}, the reflection on ethical aspects of the profession \cite{IJSHE-12}, green and sustainable software engineering \cite{SB-87}; and the environmental impact of software, networks and services \cite{SB-14}. Some studies add the sustainability of the SE process \cite{IEEE-153, IEEE-175, IEEE-172}\cite{IEEE-164}\cite{WoS-1}.

Gaining awareness can be generally mapped onto the Dublin descriptor ``Knowledge and Understanding’’. 

\textbf{Gaining practical experience} A significant number of studies propose letting the students gain practical experience in various forms: by means of integrative and transformative projects \cite{ACM-12}; by facilitating integrative learning through reflective writing \cite{SB-89}; and by exploring different ICT perspectives \cite{IEEE-153, IEEE-175, IEEE-172}.
More traditional approaches include project-centered learning \cite{SB-71}\cite{IEEE-161}, case-based learning \cite{IEEE-Penz2013cseet}, practical student assignments \cite{CompSusEd-02}; and using real-world sustainability problems, hard data and technical methods \cite{SB-92}. 

Depending on the education level covered by the study, gaining practical experience can be generally mapped on Dublin descriptors ``Applying Knowledge and Understanding’’ and ``Communication’’.

\subsection{Topics being taught (SRQ2.1)}
\label{sec:52}
The topics being taught refer to the context of computing and sustainability. In particular, we observe that teaching topics may be a collection, or even a combination, of computing-specific topics and sustainability-specific ones. Accordingly, computing and sustainability provide the context on which teaching topics are organized.

Following this line of thought, the \textit{Computing-specific topics} identified from the studies entail the phases (e.g., from the software life cycle, like requirements elicitation) and the contents (e.g., software models) specifically related to the computing field. \textit{Sustainability-specific topics}, in turn, entail notions of sustainability and in particular how it is being considered in a given educational setting (see Section~\ref{sec:4} for different conceptualisations). Finally, we have also identified a third type of somehow-hybrid topics, which stem from the combination of computing and sustainability and as such do not belong to one or the other but lie in their intersection (e.g., green computing techniques). We call these \textit{Combined Computing-Sustainability topics}. The following reports on the primary studies addressing these three types of topics.

\textbf{Computing-specific topics.} These most prominently regard the focus in a given module, specifically on software development project management \cite{ACM-10}, user experience/human-computer interaction (UX/HCI) \cite{CompSusEd-02} or measuring the energy footprint of ICT \cite{IEEE-167}. 

\textbf{Sustainability-specific topics} We  use the PESTLE classification as a general mapping of how the topic of sustainability is being understood in learning objectives.\footnote{PESTLE stands for Political, Economic, Social, Technological, Legal and Environmental. A PESTLE analysis aims to identify potential threats and weaknesses from the perspective of the six factors above. \url{https://doi.org/10.1002/joe.21935}} Accordingly, we identified studies that are specific to various notions of sustainability: the environmental \cite{IEEE-164}\cite{IEEE-167}\cite{IJSHE-5}, the social (social relevance of programming \cite{SB-53}), and the socio-technical (the impact of inclusive HCI \cite{SB-53} or social learning \cite{SB-99}). Some others are all encompassing, by addressing dimensions from the whole PESTLE classification (e.g., \cite{IJSHE-12}\cite{Scopus-1}) to sustainability in general (e.g., \cite{SB-77}). In addition, under this umbrella we identified the following topics:

\begin{itemize}
\item \textbf{Climate Change.} This content topic is addressed in many universities, some of them more in depth using Systems Thinking approaches \cite{CompSusEd-01, SB-40, SB-54}, and some of them more exploratory in projects \cite{IJSHE-21, SB-92}. The main teaching point is that climate change is present, complex, and immediate.
Krogsties and Krogsties include sustainability content in assignments, for example by means of including personas with differing views on the climate measures \cite{SB-54}.

\item \textbf{Sustainable Development Concepts and the SDGs.} Several studies used the SDGs as a source of inspiration to choose projects for students to work on, or contextualize, the application of IT in a meaningful way. Argento et al. \cite{IJSHE-26} found projects on-campus related to SDGs. Crompton et al. \cite{IJSHE-5} helped students understand how the use of technology to meet material needs of humans contributes to environmental effects. Stone [SB-8] created introductory programming projects focused on sustainability topics, such that students were exposed to the general concepts and terminology involved. This is expanded in \cite{SB-53} with reflective writing. Koniukhov and Osadcha \cite{SB-46} implemented ESD (Education for Sustainable Development, UNESCO) principles in the process of training future software engineers. 

\item \textbf{Ethics and Social Good.} Several studies mentioned the importance of including ethics and social good as context for IT projects. For example, Goldweber \cite{SB-03} reports on 14 introductory projects in that context, from radioactive mice to water pollution to voting simulation. Polmear et al. \cite{SB-15} compare macroethics in teaching across cultures that can be drawn from Hazas and Nathan \cite{SB-62} position students becoming ambassadors for change.
\end{itemize}

\textbf{Combined Computing-Sustainability topics.} Some studies report on combined contents providing competencies like green computing techniques \cite{ACM-12}, technology assessment to analyze technological developments and their consequences \cite{SB-48} and how to help minimizing the energy footprint of software solutions \cite{SB-14}; or how to work in interdisciplinary settings \cite{ACM-06} while fostering creativity \cite{IEEE-161}. Under this umbrella, we identified the following topics:

\begin{itemize}
\item \textbf{Environmental assessment, LCA and technology assessment.} These topics entail standards as well as concepts for environmental assessment (EA); life cycle assessment (LCA); design for environment (DfE); and environmental decision making \cite{SB-30}. These topics include systems and sustainability perspectives, systems performance analysis, and economic decision-making tools as well as project evaluation. Furthermore, industrial ecology contributes to the lens of Earth systems engineering and management, supplemented by LCA and material flow analysis \cite{SB-16}. Technology assessment considers how to evaluate the outcomes and impacts of large-scale complex engineering systems from a decision-maker's perspective. There are legal, political, economic, environmental and social frameworks that can be used to assess technology and its impacts \cite{SB-48}.

\item \textbf{Green IT and Sustainable Computing.} Green IT is the study and practice of designing, manufacturing, using, and disposing efficiently of computers, servers, and associated subsystems like monitors, networking and communications systems \cite{SB-29} (see also the incremental perspective in Section~\ref{sec:42}). This topic area includes energy-aware services, investments in Green IT, measurements of energy efficiency and of software qualities, and ``Green Labs" \cite{SB-04}. It can be expanded towards a circular economy, smart systems, as well as Quality of Service and environmental requirements and contracts \cite{SB-14}. 

\item \textbf{Sustainability in/by IT.} This topic area covers the  three perspectives described in Section 4.2. For Sustainability in and by IT, Hilty defines three impact dimensions of ICT (first, second, and third order effects) \cite{hilty2015ict}. These impact dimensions are used in several publications to explain the connection between ICT and sustainability \cite{SB-01}\cite{SB-05}.

The biggest potential to motivate students to engage with the topic, according to Hilty and Huber \cite{IJSHE-11}, is an understanding that ICT is part of the problem and the solution. They give examples for recycling of ICT, using ICT to reduce greenhouse gas emissions, seemingly dematerialized economies, and rebound effects that lead to increasing demand for goods or services.

Penzenstadler et al. \cite{IEEE-Penz2011CSEET} argue for making the link between IT and sustainability explicit, and offer ways of how to teach about it. \cite{ACM-06} and \cite{IEEE-Penz2013cseet} show examples of how to design the requirements for a system such that they address the impact of sustainability requirements on IT (i.e., requirement analysis). 
\end{itemize}

\subsection{Educational Organization (SRQ2.2)}
\label{sec:53}

In the following, we present ``how'' sustainability education in computing is being implemented in terms of two elements: \textit{(i)} pedagogical approaches, or the teaching methods that are used to realize the learning objectives described before, and \textit{(ii)} the ``building blocks’’ in which sustainability education is organized, i.e., modules and curricula. 

\subsubsection{Pedagogical Approaches}\label{sec:531}
We could trace two aspects in the presentation of pedagogical approaches: \textit{(i)} practices to integrate different fields of knowledge, and \textit{(ii)} pedagogical methods in sustainability education. They are described in the following.

\textbf{Practices to integrate different fields of knowledge.} Several of the primary studies include reflections on pedagogical approaches to integrate different fields of knowledge \cite{ACM-10, ACM-11, SB-15, SB-16, SB-30, SB-31, SB-33, SB-48, SB-62, SB-68, SB-82, CompSusEd-02, IEEE-171}. Multiple papers argue that the students’ discipline needs to be combined with domain knowledge of the various dimensions of sustainability \cite{SB-30, SB-16, SB-15}, and that specifically, ethics \cite{SB-30, SB-31, SB-82} or other subjects from the humanities \cite{SB-33} are central to the integration of holistic perspectives on engineering and management. Two such approaches emerged from our analysis: 

\begin{itemize}
\item \textbf{Multidisciplinary student groups.} This approach is about the composition of student groups. Multidisciplinary groups of students are advocated, allowing for a more holistic understanding of sustainability issues \cite{CompSusEd-02, SB-48}. One example is a collaboration between IT and Material \& Earth Sciences \cite{IJSHE-12}. Another one \cite{IJSHE-21} uses a question-based approach to select topics that then require looking into different disciplines.

\item \textbf{Engaging external actors.} Several of the studies report on the value of engaging external actors \cite{SB-01, SB-04, IEEE-Penz2013cseet, SB-34, IJSHE-12, CompSusEd-02, IEEE-156, SB-14, SB-79, IEEE-163}. Collaboration with external stakeholders is either reported as being done through entrepreneurial approaches \cite{IJSHE-12}, through cooperation with the industry or public sector \cite{CompSusEd-02, IEEE-156, SB-14, SB-79}, or through the local community \cite{IEEE-163, IJSHE-12}. Representatives from these sectors can give guest lectures, as e.g., industry representatives in \cite{IEEE-172}. Students become active agents of change in the local community when working with external stakeholders, as e.g., expressed in the following: “Entrepreneurial experiences is an educational framework encouraging students to have the competencies to make a positive change in their own context, the ability to assess opportunities and resources for making a difference and the courage to act accordingly”~\cite{IJSHE-12}. 
\end{itemize}

\textbf{Pedagogical methods in sustainability education.} The primary studies report on different methods, that is different ways of engaging students in learning activities and promoting skills, attitudes and behaviors assumed to be associated with learning for a sustainable development:

\begin{itemize}
\item \textbf{General learning approaches.} These are meant to help students become active participants in their modules, which are not necessarily specific to learning for sustainable development. In this category, the studies report on using student-centered active learning~\cite{SB-34}, discussions and reflections \cite{CompSusEd-02, IJSHE-8, SB-01, IEEE-172}, preparation and classification of ready-made project ideas on computing for the social good \cite{SB-03}, experiential and reflective learning activity and reflective reports \cite{ACM-11}, a “Sustainable programming pedagogical model” with reflections~\cite{SB-89}, and flipped-classroom for critical thinking.

\item \textbf{Games, simulations or role-playing exercises.} Such exercises for sustainability were described in five studies \cite{CompSusEd-01, IEEE-New-3, SB-48, SB-64, ACM-01}. Some use custom board games designed to build experiences related to understanding norms, values or complex systems \cite{SB-64, IEEE-New-3}; some use simpler playbook formats featuring smaller exercises fit for classroom use \cite{ACM-01, CompSusEd-01}; and one reports on role-playing and “structured academic controversy forum”~\cite{SB-48}. Eriksson et al. \cite{CompSusEd-01} use in their CS module the systems thinking games Living Loop and Harvest from the ``Systems Thinking Playbook for climate change''~\cite{Sweeney2010}, to let students experience the concepts detailed in the preceding lectures. Easterbrook~\cite{SB-40} reports on a seminar series where the students ``collectively tease out the systems concepts exhibited by the game, along with their own reactions to the overall behavior of the system’’.

\item \textbf{Problem-based learning, course projects and hands-on experience.} These pedagogical methods were described in fifteen of the primary studies \cite{SB-01, SB-05, SB-16, SB-33, SB-34, SB-62, SB-79, WoS-2, ACM-06, IJSHE-10, IJSHE-12, IJSHE-21, IEEE-145, IEEE-163, IEEE-172}. In these methods, the students engage with and take responsibility for real-world issues; they also take responsibility for their own learning while working on the issues. The computing students practise what is emphasised as important for engineers by Hazas and Nathan~\cite{SB-62}, i.e., taking responsibility, Corporate Social Responsibility, and communication. Mi\~{n}ano Rubio et al.~\cite{SB-33} describe course projects as an “important leverage point for the integration of sustainability”. The projects are contextualised in different ways, e.g., as a part of a longer education and vertically integrated through longer periods of studies~\cite{IJSHE-10}, or integrated with other disciplines in interdisciplinary projects \cite{ACM-06, IJSHE-21}. The concept “learning by doing” is used and extended into ``learning by living’’. The students apply and learn about software technology in specific applications, e.g., gardening (growing your own food)~\cite{ACM-06}. In one example, students changed lifestyle for a shorter period of time and reflected on the outcome, i.e., the students explore the notion of sustainable living in their everyday life, and experience first-hand the related choices and challenges~\cite{IJSHE-8}. In addition, connecting computing to social sustainability can bring sustainability to life \cite{IJSHE-6, SB-03}. Mann~\cite{SB-05} proposes an educational model for transformative action, so that students act consciously and responsibly within their professional practice. 

\subsubsection{Building Blocks: Modules and Curricula} \label{sec:532}
The realization of sustainability in computing education is discussed in terms of the changes required to modules or curricula. Some authors aspire for radical change \cite{SB-01, SB-05} and propose to rethink education as whole, as ``the solution is not to try and cram more material into an already crowded curriculum, but rather to see sustainability as the context, a basis for deeper learning, or even a reason for learning”~\cite{SB-05}, and therefore ``rather than carefully aligning sustainability with the discipline, we need to be pointing out how much change is required”~\cite{SB-05}. Teaching in ways that allow for and encourage activism, also expressed as ``development activism”, may be one approach to achieving radical change through education, for example as in ``Dissent 101”~\cite{Huish2013}. The scope of the studies varies from individual modules to the design of full curricula.

\textbf{Modules.} The primary studies report on specific modules through which sustainability education is implemented, e.g., teaching general sustainability competencies \cite{ACM-11} or cross-cutting skills like systems thinking \cite{SB-40}. The studies provide guidance for module design, for instance thesis projects about sustainable engineering~\cite{IEEE-165}. Even though many of the primary studies do not inform about the curricular level in which their proposal is being taught, most works seem to focus on the BSc and master levels. The extent to which the integration of sustainability was performed varies. In a lighter mode, we found efforts to teach sustainability in IT via one or several workshops \cite{IEEE-156} or seminars \cite{SB-87, ACM-11, ACM-12}, assignments \cite{CompSusEd-02}, projects \cite{SB-92, SB-08, SB-05, SB-33}, or through modules in modules \cite{SB-07, SB-87, SB-89, SB-53, IEEE-Penz2013cseet}. Penzenstadler et al.~\cite{IEEE-Penz2011CSEET} describe how integrating sustainability can be achieved through minor changes, e.g., in seminars, and how those minor changes can be a seed for promoting the value of sustainability to students, which might lead to more transformational changes of computing education later on. Easterbrook \cite{SB-40} informs about their determination to educate all their MSc and BSc students about sustainability and reports on how they increase sustainability-related aspects in different modules, particularly in those of their Human-Technology Interaction MSc programme. 

A few primary studies propose whole new modules such as a module on green computing \cite{ACM-12, ACM-01, CompSusEd-01, IEEE-163, WoS-2, SB-68, SB-57, SB-45, SB-29, IJSHE-5}, introductory modules on software \& sustainable development \cite{SB-75, SB-44, SB-30, SB-19}, final year project module on sustainability \cite{IEEE-165}, “software engineering sustainability” \cite{IEEE-167, IEEE-New-3}, and ethics \cite{SB-82}. Easterbrook \cite{SB-40} proposes three modules, on climate informatics, on the social and environmental impacts of the internet, and on systems thinking for global problems. 

\textbf{Curricula.} Curricula embed sustainability either as a central focus e.g., \cite{CompSusEd-03}, or as general design guidelines e.g., \cite{IEEE-147}\cite{IEEE-164}. The level of the curricula at which sustainability education in computing is implemented varies from K-12 \cite{IEEE-156}, to BSc and MSc \cite{SB-40}, to PhD degrees \cite{Scopus-1}. Several primary studies report on how sustainability is being implemented across the curriculum, or programme \cite{ACM-10, IEEE-147, IEEE-164, IEEE-171, SB-79, SB-71, SB-16, SB-33, SB-34, SB-35, SB-40}. Examples include the integration of  ``Sustainability and Social Commitment" skills (SSC) in different degree modules in a 4-year Computing Engineering degree \cite{IEEE-171, IEEE-172, IEEE-Penz2013cseet}, ``greenifying curricula” \cite{IJSHE-6, IEEE-164, WoS-1, SB-15, SB-14, SB-04, IJSHE-26}, a 1-year IS Master program “with a special focus on the Digital Society and Sustainability” \cite{CompSusEd-03}, or the blueprint of the Erasmus Mundus Joint Master Degree in Pervasive Computing and Communications for Sustainable Development \cite{IEEE-172, IEEE-175}. Mi\~{n}ano Rubio et al. \cite{SB-33} propose a model for the holistic and systematic integration of sustainability competencies into engineering curricula that can serve as guidance.

Finally, Mann et al. \cite{ACM-14} develop recommendations for policies that can catalyse the integration of sustainability in all computing programs across institutions. The vision of this work is to create “a philosophy of Computer Education for Sustainability” that “will be enhanced if undertaken within a context of institutional operational practice” (p.191).
\end{itemize}

\subsection{Effects of Sustainability Education (SRQ 2.3)}\label{sec:54}
The effects of teaching sustainability have been studied from two different perspectives: \textit{(i)} what researchers/educators reported that their students demonstrated to have learned (i.e, the achieved learning outcomes), and \textit{(ii)} other effects that were reported as a consequence of implementing their pedagogical initiatives. 

\subsubsection{The Achieved Learning Outcomes}
The achieved learning outcomes were categorised in the themes shown in Table~\ref{tab:achievedLOs} together with a description of their meanings and the primary studies in which they were witnessed. For each theme of learning outcomes, the first column (and the descriptions below) refers to the categories of learning objectives that they respond to (see horizontal pillars in Figure~\ref{fig:BigPicture}). Following the principle of constructive alignment \cite{Biggs2003}, our aim is to help educators align the \textit{intended} learning objectives and the \textit{achieved} learning outcomes. (See also the discussion in Section~\ref{sec:54}).

\textbf{Achieved cross-cutting competencies.} The primary studies report on the achieved learning outcomes that can be interpreted as cross-cutting competencies:
\begin{itemize}
    \item \textbf{Increased Awareness}. The most documented effect of sustainability education in computing is an increased awareness of what sustainability is and the sustainability issues humanity is facing. For example, Stone reveals that “a majority of participating students reported that the course and its assignments helped them have a greater understanding and appreciation for sustainability” \cite{SB-08}, an observation similarly made in \cite{SB-53,SB-69,IEEE-New-3}. Furthermore, \cite{SB-92} reports that one student “claimed not to have understood the full implications and scope of climate change before, and said that his understanding had improved much”. The students seemed surprised how much they learned and how little they knew before~\cite{IJSHE-13}. Students also enter and select sustainability education with an awareness and desire to learn and care for the environment and state they become even more aware during the education~\cite{IJSHE-5}. In \cite{SB-57} one can read how some students went from doubting climate change to expressing deep concerns for the future of mankind.
    
    \item \textbf{Restlessness.} Teaching sustainability was sometimes reported as raising on students an overall feeling of restlessness~\cite{SB-19, SB-57, SB-82}. Students expressed deep concerns, anxiety and fear about our global future. Concrete evidence suggests that being thoroughly exposed to real-world sustainability issues can lead to uncertainty on whether it will be possible to address those issues: “The revelations have made me more anxious about life and our future. It's good stuff but I would probably be happier without it. Not knowing is bliss.” \cite{SB-19}. Further evidence reports even more profound impacts, where several students have expressed feeling almost depressed halfway through the course: “The course can give a feeling of hopelessness.” \cite{SB-57}. As discussed in \cite{SB-57}, this effect suggests that the students were emotionally highly engaged in the sustainability topic.
    
    \item \textbf{Mindset Change.} In addition to an increased awareness of the topic of sustainability, the primary studies present a change in the mindset of the students as a learning effect \cite{IJSHE-13, SB-57, SB-69, SB-82, SB-92}. Issa et al.~\cite{SB-69} received formal and informal feedback from their students and concluded that their unit shifted students’ mindset. The students recognised the potential damage of IT systems and were oriented to “using IT wisely” including minimizing IT usage. In \cite{SB-82}, one can find direct feedback from students that sustainability education has changed their view of the field of Informatics and made them reason about aspects that never crossed their minds. Similarly, one student admitted to thinking about the people talking about climate change as extremists before taking the course, a vision that changed after taking it \cite{SB-92}. When asked to write a newspaper headline that would reflect any change in their thinking about sustainability as a result of attending the course, the students described in \cite{IJSHE-13} submitted headlines that reflect how confronting the course was for some of them; a concrete example being: “I am Lost!! – Please Show Me the Way to Sustainability”.
    
    \item \textbf{Increased Motivation.} Students have demonstrated an increased motivation to adopt a more sustainable practice or attitude as an effect of sustainability education \cite{ACM-12, IJSHE-13, SB-53, SB-57, SB-69}. We find expressive statements such as the “most significant [change this course has brought] has been the reawakening of an ambition to make a difference'' \cite{IJSHE-13}. The increase in motivation expressed by students also encompasses their professional practice: students who concluded a course on green computing mentioned a strong enthusiasm for green computing as something they will apply in their profession \cite{ACM-12}. Finally, one can read from a student that: “I was interested in marketing and stuff like that before the course, but now I feel like doing something that is more beneficial to humanity”~\cite{SB-57}.
    
    \item \textbf{Multidimensionality of the Problem.} One of the challenges of addressing sustainability is that it entails multiple dimensions that need to be addressed in an holistic way. One of the reported effects of teaching sustainability was that students learned the multidimensionality of the problem at hand~\cite{SB-19, SB-57, IJSHE-13}. In fact, besides understanding the multiple dimensions of sustainability, students have also shown to understand that the different dimensions are often in conflict with each other, specifically with the economic dimension.
\end{itemize}

\begin{table}[ht]
    \centering \footnotesize
    \begin{tabular}{p{1.8cm} p{1.8cm} p{7cm} p{3cm}}\toprule
        \textbf{Pillar} & \textbf{Theme} & \textbf{Meaning of the category} & \textbf{Primary studies} \\
          & (achieved learning objectives) & Adding sustainability to education \dots &  \\\midrule
          Gain awareness & Increased awareness & \dots improved the awareness among students of the global sustainability issues the world faces. & \cite{SB-07, SB-53, SB-69, IEEE-New-3, SB-92, ACM-10, ACM-11, ACM-12, IJSHE-13, IJSHE-5, SB-08, SB-57, ACM-New-2} \\
          Gain awareness, Acquire cross-cutting competencies & Restlessness & \dots led students to express deep concerns, even anxiety, about the current (lack of) sustainability of the global trends. & \cite{SB-19, SB-57, SB-82} \\
          Acquire cross-cutting competencies & Mindset Change & \dots led students to radically change their perspective about sustainability, its causes or potential solutions. & \cite{SB-69, SB-82, SB-92, IJSHE-13, SB-57, ACM-New-2, IEEE-New-3} \\
          Acquire cross-cutting competencies & Increased Motivation & \dots led students to express their willingness to adopt a more sustainable practice or attitude. & \cite{SB-53, SB-69, ACM-12, IJSHE-13, SB-08, SB-57} \\
          Acquire cross-cutting competencies & Multidimensionality of the Problem & \dots led sustainability to be recognised as a multidimensional problem, and one whose dimensions are often in conflict. & \cite{SB-19, IJSHE-13, SB-57} \\
          Acquire cross-cutting competencies & Negative Impact of ICT Systems & \dots linked the computing discipline and its impact on climate change and other sustainability challenges. Students understand how computing can harm sustainability. & \cite{SB-69, ACM-10, IJSHE-8, SB-45, IEEE-New-3} \\
          Acquire topic-specific competencies, Gain practical experience & Positive Impact of ICT Systems & \dots showed the potential of computing for addressing sustainability challenges. Students understand how computing and how their computing skills can contribute to improving sustainability. & \cite{SB-53, SB-54, SB-69, IEEE-New-3, SB-92, ACM-10, ACM-12, IJSHE-8, SB-45, ACM-New-2, IEEE-New-1} \\
          Gain practical experience & Lifestyle Change & \dots made students change their personal, or their community’s, practices and attitudes into more sustainable ones. & \cite{SB-53, IJSHE-5, IJSHE-8, SB-45} \\\bottomrule
    \end{tabular}
    \caption{The achieved learning objectives mapped on the pillars of Figure~\ref{fig:BigPicture} and general themes}
    \label{tab:achievedLOs}
\end{table}

\textbf{Topic-Specific Competencies.} The primary studies reported that the students learnt about the following topics:
\begin{itemize}
    \item \textbf{Negative Impact of ICT Systems (or the use thereof).} 
    The instructors of a Master degree course on sustainability and Green IT describe that “Students were stunned by the damages of information technology usage locally and globally” \cite{SB-69}. Similarly, having analysed the answers of the retrospective survey of a course addressing Sustainability for media technology students, authors acknowledge that they were successful in explaining the connection between ICT and climate change~\cite{SB-45}. Furthermore, in another retrospective survey, students assessed the statement “Information Technology [...] causes new problems to arise that must be addressed.” with an average of 4.62 on a 1-to-5 Likert scale~\cite{ACM-10}. Finally, in their reflections over the post-course feedback provided by students, the instructors of \cite{IJSHE-8} note that, even if reported in limited cases, students demonstrate having understood that the convenience provided by technology can lead to unsustainable habits--one clear example being online food ordering systems.
    
    \item \textbf{Positive Impact of ICT Systems.} 
    We have also found numerous examples that show that students learned about the potential positive impact of ICT systems, i.e., about the potential contributions that these can have to address sustainability issues \cite{ACM-10, ACM-12, IJSHE-8, SB-45, SB-53, SB-54, SB-69, IEEE-New-3, SB-92}. Concrete feedback from students includes “This unit is very important for future students to make the changes in the IT world" \cite{SB-69}; and that the use of climate models provided them with new perspectives on what their contribution as future engineers can be \cite{SB-92}. Similarly, we observed that students acknowledged that sustainability is a concern for both the consumers and the designers of IT systems \cite{IEEE-New-3}. The majority of respondents on a retrospective survey also agreed to strongly-agreed that thanks to the course they understand the applicability of programming to solve complex social problems \cite{SB-53}. Concurrently, students strongly agreed that “IT makes potential solutions to the problems of sustainability available [...]”~\cite{ACM-10}. The use of technology has also been described by students as a way to assist them in living more sustainably, namely by using mobile applications that track their travelling carbon footprint~\cite{IJSHE-8}.
\end{itemize}

\textbf{Practical Experience in the form of Lifestyle Change.} The primary studies report that the students learn about and test concrete lifestyle changes~\cite{IJSHE-5, IJSHE-8, SB-45, SB-53}. This could be interpreted as the cross-cutting competence “gaining practical experience”.
Interestingly, while the education was focused on the relationship between computing and sustainability, the changes most often refer to daily activities and/or community practices instead. In \cite{IJSHE-5}, one can find various concrete statements along these lines, and in particular regarding:

\begin{itemize}
    \item \textbf{Travel patterns.} “I now cycle to work”, “Cut down on unnecessary journeys”, “Have scrapped our second vehicle and use public transport or walk instead”, and even “Moved house to reduce travel to and from work”.
    
    \item \textbf{Consumption of energy, materials and food.} “We now recycle cardboard and plastic as well as glass and cans. We shop for food now with an awareness of 'food miles' and unnecessary packaging and what's in-season”, “Buy only two newspapers per week, [...], compost all relevant household waste, change all lighting to low energy”. Concurrent observations in consumption patterns can also be found in \cite{SB-45} and \cite{SB-53}. Further evidence of improved practices can also be found in \cite{IJSHE-8}, most of which reflect incremental (instead of radical) changes. We should mention that lifestyle changes are sometimes acknowledged as not being easy to achieve: \cite{IJSHE-8} describes situations in which students admitted to struggling to adjust to new lifestyle choices, but eventually persisted on them and were successful in taking (self-admitted) small steps. Finally, the majority of respondents of a survey in a study that used sustainability-themed projects in introductory programming modules have also agreed/strongly-agreed they can now see the potential application of sustainability practices in their own communities which was facilitated by the assignments and activities of the module \cite{SB-53}.
\end{itemize}

\subsubsection{Other Effects of Sustainable Computing Education} 
The effects we describe in this section are effects that we extracted from the primary studies but go beyond the intended- or achieved learning outcomes. Similarly to the previous Table \ref{tab:achievedLOs}, we extracted and classified such effects into general themes (see Table~\ref{tab:otherEffects}) described in the following. 

\textbf{Changing student demographics.} Integrating sustainability into computing education makes computing education attractive to very diverse groups of prospective students. For example, while computing disciplines have traditionally been rather male dominated, this is changing when sustainability related contents are emphasized~\cite{IEEE-175}. Stone~\cite{SB-08} reports that “the inclusion of socially-relevant projects and course themes has been shown to assist in attracting students to Computer and Information Science, and practical, problem-based applications have also been shown to attract females and underrepresented groups to the discipline”. 

\begin{table}[ht]
    \centering \footnotesize
    \begin{tabular}{p{1.4cm} p{2cm} p{6cm} p{4cm}}\toprule
        \textbf{Pillar} & \textbf{Theme} & \textbf{Meaning of the category} & \textbf{Examples of primary studies}\\
          (educational organization) & (achieved learning objectives) & Adding sustainability to education \dots &  \\\midrule
          %
          Pedagogical approaches & Changing student demographics & \dots attracts new types of students and may change the demographics of the student groups. & \cite{SB-08, IEEE-175} \\
          %
          & New career opportunities & \dots provides new opportunities in work life & \cite{SB-32, IEEE-172, IEEE-175, SB-62} \\
          %
          Modules & New proposals and openings & \dots will eventually lead to new modules, curricula and even policies. & \cite{SB-21, SB-01, SB-79, IEEE-172, IEEE-175, IEEE-171, IJSHE-10, ACM-14, SB-33, SB-87} \\\bottomrule
    \end{tabular}
    \caption{Other effects mapped on general themes}
    \label{tab:otherEffects}
\end{table}

\textbf{New career opportunities.} The integration of sustainability in computing subjects has been shown to provide students with new skills and competencies \cite{SB-71, SB-68, CompSusEd-03}. As such, they also create new career opportunities as shown in \cite{IEEE-175} and \cite{IEEE-172}. These new careers can be linked with the general evolution of society under the Sustainable Development Challenges and Climate Change in addition to required 21st century skills~\cite{CompSusEd-03}. 

\textbf{New proposals and openings.} The successful integration of sustainability into a subset of computing courses can eventually lead to new proposals and openings that may change a discipline. An example can be found in the evolution of the Educational framework for the Information Sciences masters programs, which was revised in 2016 to include sustainability~\cite{MSIS2016}. In some of the primary studies, the proposals focus on identifying elements needed for a core sustainability course (SB-01), curricula structure or content \cite{SB-33, SB-79, IJSHE-10} or even a policy/approach for including sustainability into education~\cite{SB-21, ACM-14}.

\subsection{Takeaways for RQ2}
\label{sec:55}
Building upon the results presented in Section~\ref{sec:5}, we observe that Figure~\ref{fig:BigPicture} suggests a \textit{Framework for Implementing Educational Components}. As illustrated, sustainability in computing education can be implemented by drawing on the horizontal pillars (which summarize the identified intended learning objectives), and the vertical pillars (which summarize the topics pertaining sustainability in computing, and the educational organization).


In reflection, we argue that educators could build upon Figure~\ref{fig:BigPicture} and use it as a framework to design their educational components by identifying the ingredients from each of the vertical and the horizontal pillars. For the vertical pillars, we could start defining the ingredients for the desired educational organization (Are we designing a full curriculum or a single module? Which pedagogical approach(es) do we find the most appropriate?), and then identify the specific topics we want to teach for the computing- and sustainability learning objectives (Which phase(s) do we want to cover? Which computing-, sustainability-, and combined topics? And for which notion of sustainability?). For the horizontal pillars, we could ask ourselves which generic competencies the students should acquire. Our work and the illustration could also inform an analysis of existing programs.
 
Once an educational activity is being conducted, we can assess the achieved learning outcomes by evaluating the coverage of the various pillars. We can also use the figure to reflect on the implementation of education in a larger sense, e.g., reflecting on what modules are missing or the pedagogical approaches that are lacking. Such a framework could also be an instrument to accommodate change in the way we design and run education.

%% file: 6_Results_RQ3.tex
\section{Research Design (RQ3)}
\label{sec:6}
A study’s research design provides the blueprint for the collection, measurement, and analysis of data, and aims to reduce the bias and increase the trust and reliability in the accuracy of the data collected for the research phenomenon or problem under investigation~\cite{Coe2021}. In RQ3, we investigated the research design being conducted by the community against the proposed ACM Empirical Standards~\cite{Ralph2020} guidelines with respect to the overall research design including the following sub-research questions: (i) the epistemological stance underpinning the research design, (ii) the research method(s) used and to what extent the method(s) included core attributes of the approach, and (iii) what threats to validity were considered, which potentially jeopardize the reliability of the results and the confidence in the conclusions that can be drawn.

\subsection{Epistemology (RQ 3.1) and Threats to Validity (RQ 3.3)}
Based on the results of our analysis we consider these two sub-research questions together rather than separately as a result of the brevity of the data that was identified. Sub-research question 3.1 investigated the epistemological stance adopted in the selected research literature. Epistemology is the “theory of knowledge” and influences how that knowledge is collected and from which sources” ~\cite{Creswell2018b}. In research terms, it is argued that the philosophical standpoint should be made clear from the beginning, as a researcher's view of the world and knowledge strongly influences the choice of methodology and methods, and interpretation of data~\cite{kelly2012}. Within epistemology, there are several approaches and branches, such as positivism and interpretivism, which are diametrically opposed. A thorough treatment of the relationship between ontology, epistemology, methodology, and methods is beyond the scope of this paper, and readers are referred to Bridges~\cite{Bridges2017} for a more in-depth discussion of the relationship and importance of epistemological stance and research design. During the analysis, we found no studies that [explicitly] stated an epistemological stance, which is considered a key founding principle in guiding research in education, underpinning the consequences for knowledge construction within the research community~\cite{Elmore2012}. This leads to the question of whether it is possible to derive an implicit epistemological stance through a reverse mapping between method and epistemology; however, that is beyond the scope of this already extensive work. The absence of an [explicit] epistemological stance raises important questions regarding research design overall, and the reliability and generalizability of the results as it is concerned with all aspects of the validity, scope, and methods of acquiring knowledge. In addition, it influences how researchers frame their research in their attempts to discover knowledge and the extent to which its transferability can be assessed. 

Sub-research question SRQ3.3 investigated the threats to validity in the identified research literature. Threats to validity refer to specific reasons as to why researchers can draw incorrect conclusions when they make an inference in an experiment because of covariance, causation constructs, or whether the causal relationship holds over variations in persons, setting, treatments, and outcomes~\cite{Creswell2020}. While the concept of threats to validity has evolved over the years from the initial discussions by Campbell and Stanley~\cite{Campbell1963}, Shadish, Cook, and Campbell~\cite{Shadish2002} identified the potential threats to validity, and the statistical procedures of educational experiments, that are generally considered in modern-day research including statistical conclusion validity, construct validity, internal validity, and external validity. From an educational research perspective, Creswell~\cite{Creswell2018a} considers two primary threats critical to consider: internal validity and external validity, where internal threats relate to drawing appropriate inferences to the actual design and procedures used in an experiment, while threats to external validity are problems that threaten a researcher's ability to draw correct inferences from the sample data to other persons, settings, treatment variables, and measures. According to Cook and Campbell~\cite{Cook1979}, three threats may affect this generalizability: the interaction of selection and treatment, the interaction of setting and treatment, and the interaction of history and treatment. During the analysis, we found N studies that explicitly considered threats to validity in their research design, which are considered key characteristics of experimental research in education~\cite{Creswell2020}. The general absence of threats to validity raises important questions regarding research design overall and the reliability and generalizability of the results.

\subsection{Methods (RQ 3.2)}
Sub-research question 3.2 investigated the methods adopted in the primary studies and to what extent they aligned with the essential attributes characterizing each type of empirical study. Table \ref{tab:RQ32Table} shows the number of studies that belonged to each type of empirical study and the number of essential attributes we used to evaluate the quality of each paper. We used the classification from the ACM~\cite{Ralph2020} to categorize the type of empirical studies such as shown in Table~\ref{tab:RQ32Table}. 

\begin{table}
\begin{tabular}{c c c c c c}

\toprule
Type of Empirical Study & Number of Studies & Number of Essential Attributes \\
\midrule
General standard & 59 & 16 \\ 
Systematic review & 2 & 13 \\  
Questionnaire survey & 11 & 18 \\
Interview study & 5 & 10 \\
Case study & 13 & 12 \\
\bottomrule

\end{tabular}
\caption{Number and type of studies classified according to ACM Empirical Standard~\cite{Ralph2020} }
\label{tab:RQ32Table}
\end{table}

\textbf{General standard} This general standard applies to all studies that collect and analyze data. The results of our analysis identified that the majority of the studies were classified into the General standard category as reported in Table \ref{tab:RQ32Table}. Focusing on the research type of the papers within the category, we classified them into the following sub-categories: (i) a vision/proposal paper describing an educational plan for the future, (ii) a position paper recommending a course of action, (iii) an experience report, and (iv) a proposal of something already happening. From our analysis, we found that the majority of the studies classified in this General category can be considered “Experience reports” \cite{ACM-11, ACM-12, CompSusEd-01, IEEE-153, IEEE-New-2} or “Proposals” \cite{CompSusEd-02, IEEE-156, IEEE-147}.
A small number of papers were classified as “Vision” papers, such as \cite{IEEE-161, SB-19}. With regards to the attributes that describe the purpose and objective of the research questions we found some diversity. For example, in the experience papers the studies report that the most common problems belong to the following topics: (i) study the impact of sustainability software engineering education \cite{ACM-11, IEEE-175, IJSHE-10}, (ii) develop green computer courses \cite{ACM-12, IEEE-163, SB-04}, (iii) studies suggesting system thinking games, creativity and practical assignments to introduce sustainability~\cite{CompSusEd-01, IEEE-Penz2013cseet}, and (iv) curriculum design to introduce sustainability competencies in software engineering or computer science courses \cite{SB-21, SB-34}. For proposal papers, we also found similar topics to that of the experience report papers such as assignments related to various sustainability topics (\cite{CompSusEd-02}), curriculum design and competencies (\cite{IEEE-147}) as well as proposals suggesting how to teach sustainability and social skills (\cite{IEEE-171}), green computing models for sustainable development \cite{IEEE-164, SB-04, SB-29}, and introducing sustainability into bachelor courses \cite{IEEE-Penz2011CSEET, SB-01, WOS-5}. The vision paper~\cite{SB-19} discussed future ways to teach sustainability based on an introductory course in sustainability and ICT, while another vision paper~\cite{WoS-new-1} describes an educational plan of sustainability in computing for the future.

Another essential attribute we analyzed was if the methodology was appropriate for the goal or problem stated. A significant number of studies did not clearly report the methods used for experience report papers~\cite{CompSusEd-01, SB-21} or were “not applicable” in the case of position papers~\cite{IEEE-147, IEEE-153}. The reason is that we noted that some vision or proposal papers under this category did not carry out any research, and consequently including a research method does not make sense. Other papers partially report the methods used somewhere in the paper \cite{ACM-11, ACM-12} but there is no explicit methodology section that describes the methods employed for data collection and analysis. Finally, there are a set of papers that clearly include a clear methodology section used in the studies \cite{IEEE-Penz2013cseet, IJSHE-10, IJSHE-8, SB-14}.

With regards to the what, where, when, and how data were collected, only some studies in this category indicated fully or partially how and where the data was collected \cite{ACM-11, ACM-12, CompSusEd-01, CompSusEd-03, SB-14, SB-21}, while the others did not provide such information or was “not applicable” according to the study. Also, many of the studies provided the number of subjects that participated in the experiments or surveys ranging from few number of subjects (e.g. 16 in \cite{SB-14} or 20 in \cite{ACM-11}) to a huge number participants (e.g. 1270 in \cite{IEEE-175} or 2420 in \cite{IEEE-164}). Other studies did not indicate the number of subjects. Surprisingly, for the rest of the 11 essential attributes defined for this category (e.g. discuss threats to validity, clarify the role of the researchers, innovates the research methodology or discuss statistical power or saturation), in the majority of the papers analyzed we could not find this information or this attribute was “not applicable” based on the type of study being reported, e.g. experience. 

\textbf{Systematic review:} In this category, we defined 13 essential attributes but we only found two studies \cite{IEEE-145, SB-32}. \cite{IEEE-145} study is a systematic literature review (SLR) that investigates research trends on sustainable development for training software engineers on sustainable principles, while \cite{SB-32} is a survey that provides an educational landscape of green ICT and sustainable computing. Both studies partially describe the problem objective or research question and both outline an appropriate methodology. Only~\cite{SB-32} provides the information regarding data collection. The rest of the essential attributes defined in the ACM standard have not been included in \cite{IEEE-145}. However, in the case of~\cite{SB-32} the authors mention the detailed steps and replication package of the process, the inclusion and exclusion criteria, and the data extracted from the primary studies.

\textbf{Questionnaire survey}: In this category we classified eight studies where open-ended questions are systematically analyzed. Several types of goals are investigated. For instance, in~\cite{SB-15} the authors mix a questionnaire survey with qualitative questions about ethical aspects that include sustainability. Ethical aspects are also investigated in~\cite{IEEE-170} to illustrate how sustainability practices include ethics into teaching. The study uses open-ended questions to extract additional thoughts from the participants. In~\cite{ACM-14} a survey compares the understandings of sustainability using different benchmarks and discusses the attitudes of the participants towards integrating sustainability in Computing. Another work~\cite{ACM-New-2} investigates the use of green computing techniques and how green IT topics are taught in CS courses. About the rest of the essential attributes that characterize this category, only \cite{ACM-14} did not include the research questions while~\cite{ACM-07, IEEE-170, ACM-14, SB-15, SB-31} included these. Additionally, all of the studies indicated an appropriate research methodology, described how and where the data was collected and how the data was analyzed (except~\cite{ACM-14}). We did not find evidence for the rest of the essential attributes (e.g. how responses were monitored or managed, analysis of response rates, or measuring the constructs using validated scales). Only some of the studies partially indicated that, for instance, the threats to validity~\cite{SB-15}, described systematically the replication of the study~\cite{ACM-14}, or described how the participants were recruited~\cite{IEEE-170}. Only two studies \cite{ACM-14, ACM-07} detailed how the questionnaire instrument was created.

\textbf{Interview study:} In this category we only found two studies \cite{IJSHE-6, SB-07}. The first study reports semi-structured interviews from programme leaders and teachers about how social sustainability can be integrated and taught, while the second combines quantitative and qualitative analysis to explore the awareness and conceptual understanding of environmental issues in students enrolled in an introductory computing course. The interview study investigated how social sustainability is operationalized in engineering programs and the resources needed to support social sustainability educators as a learning objective. In this second study, two groups of 137 CS/IT students were asked about 20 questions on student’s computer use and green computing. Both studies indicated the research questions and the methodology while partially indicated the details of the data analysis process and the selection process of the interviewees. Only~\cite{SB-07} described the demographics of the participants while \cite{IJSHE-6} included the experience of the subjects. Finally, none of the studies discussed how saturation was achieved or included the threats to validity. Notwithstanding, in \cite{SB-07} we found some discussion around the possible biases reflected by the researchers.

\textbf{Case study:} In this last category we found eleven studies that cover different topics. For instance, \cite{SB-08} is a two-year case study aimed to create introductory courses focused on sustainability topics, while \cite{SB-16} uses a tool to assess universities' curricula in sustainability development and two surveys were conducted to investigate the results of the assessment of sustainability courses in environmental engineering curriculum at a university in the US and the students perceptions of the contribution of such courses to sustainability. Other studies analyze the effect of teaching sustainability topics in regular courses, such as \cite{SB-30} which reports the results of a sustainability course taught along 3 years and the effectiveness for teaching sustainability, while \cite{SB-33} analyzes the learning guidelines (i.e. a document that describes the course details) of degree programs in informatics and industrial engineering in 25 Spanish universities and proposes a curriculum model to embed sustainability into engineering education. Other studies like \cite{IEEE-New-1} analyze 45 project proposals to identify social and technical sustainability aspects and how sustainability topics are introduced in customer’s project descriptions. 

From our analysis of the rest of the essential attributes characterizing case studies, we found that 3 out of 11 studies \cite{IJSHE-11, IJSHE-13, SB-08} included the research questions but none of the studies provided examples for the questions investigated. Conversely, all the studies indicate, fully or partially, a proper research method for each case study and all except two \cite{ACM-10, SB-30} indicated how and where the data was collected. Compared to the other categories, the studies indicated in most cases that the rest of the essential attributes were covered or partially covered excepting some case studies that did not include some of these attributes. For instance, studies like \cite{ACM-06} did not include the demographics of the participants and study \cite{IJSHE-11} did not describe the site in rich detail. Others such as \cite{SB-30} and \cite{SB-33} did not justify the selection of the case study while only three studies \cite{ACM-06, IJSHE-11, SB-08} included some description for the threats to validity. For the rest of the studies, most of the essential attributes were detailed. 

\textbf{Summary:} Less than half of the papers we analyzed actually present empirical research. It is challenging to summarize the quality of the studies for each category and for many different essential attributes that should characterize each of the studies. Apart from a few papers that could belong to two different categories, we found some common patterns in our analysis revealing the quality of the studies analyzed. There are a significant number of studies that, according to ACM guidelines~\cite{Ralph2020}, do not describe the majority of the essential attributes. This is especially bad in the General standard and Questionnaire survey categories. However, we cannot provide significant conclusions for Systematic reviews and Interview studies as we only found a small number of papers in these two categories. On the other hand, the Case study category exhibits good results for the papers classified. We found most of them describe the majority of the essential attributes so it seems more easy to classify papers under this category than for the others. We statistically summarize in Figure 6.1 our findings attending to the quality of the studies and the number of essential attributes supported and not supported. It should be noted that we did not include in the count the attributes labeled as “not applicable”. Also, in the left side of Figure 6.1 we show the ratio of attributes supported per study type with respect to the number of papers for each category.  

\begin{figure}
    \centering
    \includegraphics{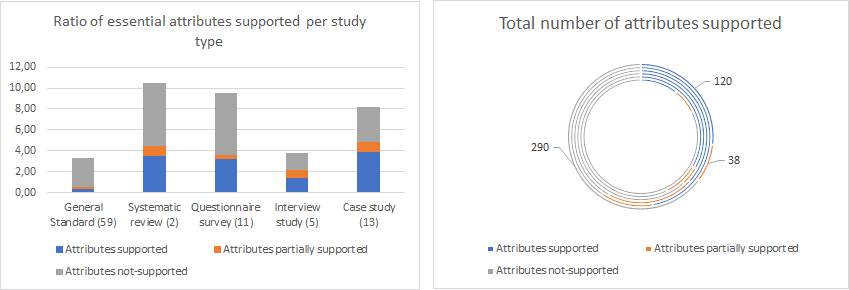}
    \caption{The number of essential attributes covered for the different types of research methods.}
    \label{fig:essential-attributes}
\end{figure}

\subsection{Takeaway Lessons (RQ3)}

This section provided an analysis of the types of studies that have been conducted with an emphasis of the specific design of the research including the procedures involved in them, such as sampling, data collection instruments and protocols, etc. In the absence of any generally agreed standard for evaluating the quality of the research design, we classified each study using the ACM Empirical Standards~\cite{Ralph2020} as a framework to aid in the identification of the essential, desirable, and extraordinary attributes for empirical research for a particular method. Alternative perspectives on evaluating the quality of research have been proposed by Lincoln~\cite{Lincoln1995}, Tuckman~\cite{Tuckman1999}, Richardson~\cite{Richardson2000}, and Creswell and Miller~\cite{Creswell2000}. In combination, they offer philosophical, procedural, and reflexive criteria to use in evaluating research design. However, Creswell~\cite{Creswell2014} argues that as a result of the fundamental differences between methodological approaches to research, each merits its own criteria for evaluation. Because of diversity in the various approaches and subject areas, there is no single Gold standard for determining the quality of research design but there is a clear need for the research community in this area to consider moving towards adopting or developing a common standard by which research output can be transparently and consistently evaluated against. Our analysis highlights that the majority of studies published are experience reports that incorporate a range of different methods. These studies are underpinned by a pragmatic and ad-hoc approach rather than a theoretically pure and rigorous approach to research design. As a result, the published output exhibits a similar alignment of shortcomings in research design to those reported by Hall, Ward, and Corner~\cite{Hall1988}.
While the papers do not follow a rigorous approach to the design of the research from a theoretical perspective, the results provide an insight into how sustainability can be integrated into the computing curricula at different levels of granularity. As a result, it is strongly recommended that researchers investigating sustainability in computing education adopt a more meticulous and repeatable approach to the quality of the research being conducted. This is imperative if any improvements or recommendations on educational practice, the quality of policy debates on educational provision, and the advancement of knowledge in the field are to be achieved that are supported by sound empirical evidence. To improve the overall quality of the research design, and the validity and reliability in data-gathering procedures, we strongly recommend: clearly stating an explicit epistemological stance; adopting a research design appropriate to address the problem under investigation; clearly describing the data-gathering methods and adopting an appropriate sampling strategy where applicable; using appropriate methods to analyse data; explicitly stating assumptions; and framing the limitations of the study around threats to validity.

%% file: 7_Discussion.tex
\section{Discussion}\label{sec:7}
This study provides understanding of existing work on sustainability education in computing and its limitations. The review by Pollock et al. from 2019 \cite{pollock2019} suggests that sustainability or climate change has been incorporated into computing education primarily at an ``aspirational level, without many tangible results to date” (p. 16).  In the present systematic review, spanning a wider set of digital libraries, we analyzed 90 primary studies (compared to 11 papers in \cite{pollock2019}). Together, the articles present different perspectives on how sustainability education can be, or already is, implemented and researched on different scales (seminars, courses, programmes) and on different levels (K-12 to PhD education).

\subsection{Limitations of existing work and directions for future work}
Most of the existing literature does not address the \textbf{severity of challenges and demand for unprecedented change}. Stein et al. \cite{stein2022:end-of-world} argue that education needs to be disruptive, it needs to be about ``growing up” and showing up differently, facing the ``impossibility of sustaining our contemporary modern-colonial habits of being, which are underwritten by racial, colonial, and ecological violence” (p. 275). We did identify work that engages with conceptions of computing as disruptive (Section \ref{sec:42}) and have also pointed to studies that argue for radical changes to computing education. As discussed in \cite{SB-45}, approaches for curriculum change vary from the incremental and integrated to the transformative and critical. Such transformative changes rely on the idea that promoting a just and sustainable society must be the goal, not computing itself \cite{sterling2004higher}. We find that the majority of studies present smaller and incremental changes to a curriculum to include sustainability rather than a fundamental redesign of education (Section \ref{sec:53}). We also found that existing work is anthropocentric, motivating sustainability for human well-being (Section \ref{sec:4.1}). 

With the exception of a few articles, \textbf{education is mostly conceived of as training} pre-defined knowledge, skills or competencies (Section \ref{sec:4.3}). We have identified an alternative understanding of education, which we termed ``emancipatory education”, inspired by previous research. Viewing education as training has been argued to be too limited. It renders education into an instrument to bring about certain competencies that are decided upon by the few, in the current unsustainable system. It locks education to being about training certain competencies thereby reducing the possibilities for alternative futures to emerge in education and limiting the potential of education for change \cite{holfelder2019towards,facer2017noveltyHope}. The potential of education for alternative futures and radical change has been argued to lie in being a space for reflection, experimentation and creativity, allowing for the possibility of the impossible, not yet imagined \cite{osberg2010TakingCareOfTheFuture,facer2017noveltyHope}. Education could be a place in which new ideas, understandings and ways of being in the world evolve, drawing on and taking into account the diverse experiences that all involved parties may bring (students, teachers, others outside school or universities) \cite{osberg2010TakingCareOfTheFuture,holfelder2019towards}. It would require to make space for the individual learner’s experiences and meanings. Such education has the purpose of ``subjectification”  \cite{biesta2009subjectificationArticle}. Ideas of education for radical change and sustainability stand in contrast to dominant framings of contemporary education. Education has increasingly become oriented towards producing certain disciplinary outcomes in an effective way and has become instrumentalised to achieve economic goals \cite{amsler2017contesting,Osberg2020,mendickPeters22purpose}. In education for radical change, concrete learning objectives may still be important, but they may be used in new ways, exploring change and potentials of education as a part of transforming society. Importantly, learning outcomes should not only be defined on the grounds of pre-existing occupational profiles and businesses or economic agendas, but rather help explore novel and potentially disruptive profiles. 

Among the work that embraces the need for radical change, we find the argument that education should play a more activistic role considering the change that is needed \cite{SB-05}. \textbf{Activism in and through education} in computing education is hardly discussed. In light of the literature on the role of education, we should be cautious not to turn education into an instrument for activism as it is perceived by certain people. This would go against the idea of education as emancipatory, non-normative and less instrumental. However, education might involve activism or direct actions as means for learning and acquiring competences. If students are to take part of making and unmaking futures, they need to be emancipated. Students need to be given opportunities for deep reflections on the root causes of the current crisis in order to contribute to a social and political movement for a fundamentally different society \cite{kaufmann2019building}. As a part of this, students need not only cognitive but also action capabilities, they need to engage with values and emotions.

Engaging with our predicaments likely evokes unpleasant \textbf{emotions} \cite{ojala_hope_2017}. Emotions have gained increasing attention in the past years in the education for sustainability literature, albeit less so in computing education. However, the learning outcomes identified in this work point to emotions: students become more aware, they become ``deeply concerned”, anxious and restless (Section \ref{sec:54}). Existing research suggests that young people today are already very concerned and even pessimistic about the state of the world. The described outcomes of education may therefore not necessarily be the effect of the education but may describe the students’ general state. An important question therefore is how education should engage with emotions. A recent review by Pihkala \cite{pihkala_eco-anxiety_2020} collects ways to address climate anxiety in education. It has inspired a recent intervention study in a sustainability course for media technology students \cite{eriksson_addressing_2022}. More than addressing negative emotions, education should also promote ``critical hope” \cite{ojala_hope_2017} or ``active hope” \cite{macy_active_2022}. Ojala \cite{ojala_hope_2017} reviews the role of education for promoting hope and points out that hope is a contested concept as it can be based on denial or de-emphasising the severity of sustainability issues. Hope, however, is argued to be an “existential must” or an ontological necessity \cite{freire1992}. Gaining critical awareness and learning about ways of regulating emotions and distress should therefore be an important part of education for sustainability. As humans we tend to only take action when motivated, for example if discomfort arises. It is therefore important for students in computing to engage and deal with emotions. We pose the open question of how such affective engagement can become stronger, considering dominant norms and values in computing that suggest computing is a rational, value-neutral endeavor.

Research on \textbf{norms, values or the culture of computing education} has been conducted for decades with a focus on understanding under-representation, mostly of women (see e.g. \cite{Rasmussen1991,margolis_unlocking_2002,salminen-karlsson_problem_2011,peters_learning_2017,peters_students_2018,ottemo_kon_2015,ottemo_contextualizing_2020}). Much of the work has been conducted by gender researchers. The research repeatedly concludes computing is socially (re-)produced or positioned as purely rational, technical or non-human, abstract and reductionist. Reductionism is not only understood in terms of reducing complex problems to smaller parts, but also avoiding to engage with problems in their complexity, approaching all problems with mathematical and technical methods. Even though rationality is valued high, emotional engagement does exist, though it is found to be directed to machines and solving technical problems. Computing education provides few opportunities to learn about technology in meaningful contexts  (e.g. computing for sustainability). Students dis-engage and adapt an instrumental approach to studying computing \cite{ottemo_kon_2015}. This culture is argued to be aligned to, or ``co-produced with”, (hegemonic) masculinity. It excludes women, puts them off \cite{salminen-karlsson2003}. This body on computing culture could inform the work on sustainability education in the future. A recent review of work on climate mitigation \cite{stoddard_three_2021} has drawn a similar conclusion: an important step is to connect the work on equality and climate change, which has so far hardly happened. They find that reasons for our collective failures to address climate change are a “narrow-techno-economic mindset” and failures of education to open vistas for imagining fair and sustainable societies.  

There are some discussions of norms and values in the articles that could gain from the connection to existing research on equality or gender. For example, Easterbrook \cite{SB-40} criticized computational thinking as the focus in computing education and suggests systems thinking as a necessary competence to work for sustainability (see also Section \ref{sec:51}, learning objectives). Computational thinking is reductionist and the work on gender and equality helps in understanding why and how reductionist framings are maintained. Another example is the work in which Mann argues that students should become change agents promoting strong sustainability, which ``will be at odds with other, less forceful sustainability perspectives”  \cite{SB-19}. He suggests: ``In the best of all possible worlds, we would like them to act as insiders who are part of the dominant culture, but who at the same time try to change the system they are part of from within.” \cite[p.19]{SB-19}. There is lots of gender research on cultures of computing, science, and engineering, as well as education in these fields that help understanding culture and how they are maintained. Connecting those two fields could be useful to understand the complex challenges and develop education for system change.

In light of the dominant framings of sustainability education in computing as training in separate modules, we raise concerns about \textbf{student well-being}. What happens to students if they learn about the great challenges we are facing in a module on sustainability, with the outcomes of being restless and very concerned, feeling discomforting emotions? They may have gotten support to deal with emotions during the sustainability module but might need to suppress anxiety or grief in the rest of education, as it lies outside the norm. Suppressing uncomfortable emotions is one unsustainable way of handling them, which can lead to depression and stress \cite{pihkala_eco-anxiety_2020}. This is another reason why sustainability cannot be left to a course or few shorter modules only. Addressing these concerns, emotions and well-being is important for all courses. There is evidence that computer workers and their products can significantly benefit from well-being practices \cite{penzenstadler2021take} that can be taught to computing students. If computing is to contribute to the well-being of others, humans and other species or the planet, then the discipline needs to evolve as a discipline going beyond the understanding of technology as a purely rational means of interaction. 
 
In light of all this, we ask the question of how to proceed with \textbf{curricula and pedagogical approaches} for sustainability education in computing. On the one hand, we might have reached a critical mass of work that could be used to define standards and curricula. It could increase the recognition of sustainability education in computing and hence promote change of norms. It may also help to push or even ``enforce” sustainability education, making sure all students get a chance to learn about and engage with different aspects of sustainability. The learning objectives described in the literature reviewed could inform such a step. On the other hand, we have identified that more encompassing views of education, which we chose to call ``emancipatory education” in-line with the literature on education for sustainability (Section \ref{sec:4.3}), are mostly absent from the material reviewed. This implies we should develop education in other ways that grant freedom and trust, allowing creativity and diverse individuals to develop and contribute to knowledge and change in ``their ways”.  

We find few publications on \textbf{more specific topics within computing and sustainability}, e.g. on software architecture or software testing and sustainability. This may be due to our search query that looked for papers within computing and sustainability. Our SLR could hence be extended to cover these computing-specific education papers. 

In order to find other holes and get a more “complete” picture of what sustainability education in computing can entail, our \textbf{findings could be mapped to other existing standards}, e.g. to the Dublin Descriptors like, for instance ``making judgments”, which emerged as especially important in sustainability in computing, for its intrinsic need of an holistic discourse. If such mapping would be present when combining computing- and sustainability topics in the design of new courses (or re-design of existing ones), educators could use the mapping to reflect on which learning objectives (e.g. “making judgements”) should be achieved with a true combination of computing-sustainability topics, like architecture assessment methods that embed sustainability dimensions, and which would instead be achieved by separate sustainability- and computing-specific topics that are orthogonal and strengthen one another (like architecture quality assessment methods on the one side, and sustainability theories on the other side). They are the European Union standard classification used in educational frameworks for curricula development and accreditation. As such, they should serve well for the purpose of integrating sustainability into computing curricula. They have however received criticism for not addressing attitudes and competencies as for example proposed by~\cite{frezza_modelling_2018}. Also, to cover the relationship between existing computing topics and what is generally understood about the societal transformations needed may be missing the point, as the changes needed are claimed to be emancipatory or at least concern new key competencies. Key sustainability competencies were only found based on the work on by Wiek et al. from 2011 (see Section \ref{sec:51}). However, these have been updated since \cite{pacis2020key}. 

We hope that this work will lead to further \textbf{developments of existing frameworks for computing curricula}. A first scan of the ACM/IEEE curriculum published in 2020 \cite{acm_ieee_curricula} shows, for example, that indirect impacts or rebound effects of computing systems are not mentioned. 
 
One question for the development of computing curricula is how to \textbf{make space for sustainability education}. We see that existing computing curricula such as the ACM/IEEE curriculum \cite{acm_ieee_curricula} get expanded item by item. Sustainability is one of many items being added or injected into the curriculum, just as other items. It is next to impossible to discard any items because they all still seem relevant to some degree by some. As argued by Klotz \cite{Klotz2021}, humans have a tendency to solve challenges by adding things and making things more complex --- rarely do we manage to instead simplify and subtract. Computing is a young discipline, so it should be easier, compared to other subjects, to fundamentally rethink the curriculum. Getting rid of mandatory curricula rather than creating ever more might be a preferable approach to engage with sustainability, especially as “sustainability add-ons” risk being seen as less relevant to core technical topics as long as a techno-centric computing culture persists. Such an idea is well-aligned with the view of education as emancipatory. It would require a broad and critical introduction to technology and our current predicaments, a change of computing culture that orients students to abstract, reductionist technical competence.
Embracing the idea of education as emancipatory (see Section~\ref{sec:53}), as something that goes beyond training that which we already know, implies we open up education and leave space for concerns and development that is not determined by the educator or other people in power. This could mean to leave spaces in the curriculum for students to define, or even moving away from the idea to design education based on curricula. 
 
Our analysis reveals that a large number of the articles are \textbf{experience reports rather than thorough research studies}. The current state of research on sustainability education in computing reminds of the state of computing education in 2004. Most of the papers then were classified as “Marco-Polo papers” \cite{valentine_cs_2004}, work of the nature “I went there [e.g. programming education, here computing education for sustainability] and I saw this” [e.g. the students liked it]. Such work is of value for sharing experiences among educators. Conducting more rigorous research could possibly increase the value of the research by providing more trustworthy evidence. Setting up standards for research could help but also limit creativity and experimentation. As argued before when discussing curricula for education, standards for research could lock research to our contemporary system and ideas of how to conduct relevant research. In any case, more empirical research is needed, using a variety of empirical research methods. Funding for such research is necessary. Section~\ref{sec:6} provides quality attributes from the ACM standard. For the experience report, the standard is less applicable so developing standards might be useful. Inspirations could also be gained from the larger field of computing education \cite{fincher_robins_2019} or from research methods used in education \cite{Cohen17researchMethodsEdu}. A new standard for research on educational experiments and interventions that assures quality and grants freedom would be valuable, i.e. guidelines for experience reports.
 
With an understanding of education as emancipatory (see Section~\ref{sec:4.3}), \textbf{the role of research} could be to \textbf{capture personal growth or novel, more sustainable ways of living, understanding, or relating to the world}. Research here might be more agile, adapting methods and theories to what emerges from education practices. Creative empirical methods (e.g. \cite{Kara20researchMethodsCreative}) might also be a source of inspiration. Another step forward could be the collaboration with like-minded people from other disciplines and external stakeholders, which is already used as a pedagogical approach (Section~\ref{sec:53}). As many have argued, we need to cross disciplines and benefit from interrelationships, rather than sticking to silo thinking.

\subsection{Threats to validity} \label{sec:ttv}


For this section we considered various classic sources, e.g.,~\cite{Kitchenham2002,wohlin2012experimentation}; however, as observed by Ampatzoglou et al.~\cite{TTV-SLR-2019}, there is no standard way of writing about the threats to validity, in general and specifically for SLRs, but it is common to include: construct validity, reliability, internal validity and external validity, according to the guideline proposed by \cite{Runeson2008,Zhou2016}, which we adopted in the following. 

\textbf{Construct validity:} 
This concerns the correct data collection and the correct measurement of the theoretical concepts \cite{Easterbrook2008}. The authors are all experts in the field of the study. While their background knowledge was used to shape the study design (and this might have introduced strong biases), we both defined an explicit protocol following the customary guidelines in the field, and refined it in a number of iterations to ensure a thorough research process. 

\textbf{Reliability:} This concerns whether the study is replicable and extensible by other researchers. We carefully documented the overall structure of our study in an explicit protocol, including the research questions, criteria for inclusion/exclusion, the search string, and the snowball process. In particular, inclusion and exclusion of primary studies are essential components of our research protocol. Therefore, we assigned two researchers to look at each paper; each gave a relevance score independently. In the case of disagreement, a third researcher looked at the paper so that consensus could be achieved. Finally, the whole group did the snowballing. For each paper, one group member identified the relevant papers. Then, a different member analysed the accepted papers, validating their relevance according to the inclusion and exclusion criteria. Again, in case of disagreement, a third member was involved.

\textbf{Internal validity:} This concerns whether researchers interpret the data correctly. Accordingly, one possible threat is misinterpretations of terms used in papers that were collected. To mitigate it, we discussed each finding in groups of at least three researchers and had weekly meetings for discussion during the past two years. We have a strong background in this domain. The researchers come from five different European countries and nine different universities. We have different experiences within the fields of sustainability, computing and education. This background gives us diverse perspectives thanks to our international origin.

\textbf{External validity:} This concerns how well the study is generalizable. Since SLRs collect papers based on a search string, any papers that meet the search string are included; bias can pertain the search string itself and the snowball activity. We discussed the search string among all authors to mitigate this threat. After several interactions of rectifying to include as many relevant papers as possible and avoid biases like culture and sex, we have obtained the status quo regarding sustainability education in software engineering. We also had a few papers we knew should be found by the search. This was one of the reasons we also included the education workshop ICT4S since that contained additional paper that we did not find in the four databases and was relevant for our study. It is possible that we have overseen papers in the snowball process, given that it was done manually. However, we tried to avoid this from happening as much as possible by having a rigours process.  We all participated in this process that lasted 22 months, where we met regularly every week to work together and discuss our findings.

%% file: 8_Conclusion.tex
\section{Conclusion}\label{sec:8}
This paper presents the state of the art on conceptions, implementations, and research on sustainability education in computing. Educators and researchers can use them to position their own work, e.g. on how they conceive of sustainability as well as computing and education for sustainability. The framework presented in Section \ref{sec:5} captures different aspects of current implementations, such as learning objectives and outcomes as well as pedagogical approaches. We find some evidence that sustainability education is appreciated by the students, they become aware, concerned, and gain competencies. We were able to map observed learning outcomes to the identified learning objectives. We have also described other outcomes, suggesting that sustainability education is about more than learning but change, e.g. of computing and education practices.

We find 90 articles describing sustainability education in computing, but the work that engages with the severity and complexity of challenges and the dysfunctioning and violence of our current system or ways of organising life is rather scarce. Adding sustainability as a topic as other topics is inappropriate. We need to transform education and explore the potential of education and computing in a process in which new practices and processes evolve. Developing sustainability education in computing needs to take inspiration from various fields of knowledge, especially critical studies of technology and sustainability education. Education should not be reduced to acquiring competence but can be seen to be about contributing to democratic change and learning as a part of the process. 

A condition for an education that takes responsibility is an affirmative or caring orientation to the world and the future. This would mean to challenge and change norms and values in computing education that encourage individuals to engage in computing as a purely technical, abstract and rational undertaking. Though understanding technical underpinnings and development trajectories of computing are relevant, reproducing or continuing on them might not be. We are missing connection to previous work on gender and under-representation that help to understand how norms and values are reproduced in power structures. Arguing for the potential of education for novelty and hope, we pose the open question of how computing education can orient itself more towards emancipation, based on care, experimentation, critical thinking and reflection.

Our times are characterized by trouble, and equally “by ingenuity and exploration, by invention and reinvention of old ideas” \cite{facer_storytelling_2019}. We are living in exciting times \cite{nicholas21_howToBeHuman}, in which we must accomplish unprecedented change, setting stop not only to emissions but to violence that is built into our modern ways of being. Sustainability education still is somewhat of a niche activity within computing education, which urgently needs to change. Knowledge and experience has been published and exists in all of us, educators and students, and needs to grow in collaboration with communities and organizations outside the university.

%% file: table-primary-studies.tex

\begin{longtable}[9]{| c | c | c | c | c |c |c |c |c |}
 \caption{List of Primary Studies: by publication year, with Venue types ([J]ournal, [C]onference, [W]orkshop, or [O]ther formats) and the main Venue Focus ([E]ducation, [C]omputing, [S]ustainability) \label{long}}\\

 \toprule
 & & \multicolumn{4}{c|}{Type of Venue} & \multicolumn{3}{c|}{Venue Focus}\\
 \hline
 Ref. & Year & J & C & W & O & E & C & S\\
 \hline
 \endfirsthead

 \hline
 \multicolumn{9}{|c|}{Continuation of Table \ref{long}}\\
 \hline
 Ref. & Year & J & C & W & O & E & C & S\\
 \hline
 \endhead

 \hline
 \endfoot

 \hline
 \endlastfoot
 
\cite{IJSHE-5} & 2002 & $\blacksquare$ & $\square$ & $\square$ & $\square$ & $\blacksquare$ & $\square$ & $\square$\\
\cite{SB-30} & 2007 & $\blacksquare$ & $\square$ & $\square$ & $\square$ & $\blacksquare$ & $\square$ & $\square$\\
\cite{SB-34} & 2007 & $\blacksquare$ & $\square$ & $\square$ & $\square$ & $\blacksquare$ & $\square$ & $\square$\\
\cite{SB-48} & 2007 & $\blacksquare$ & $\square$ & $\square$ & $\square$ & $\blacksquare$ & $\square$ & $\square$\\
\cite{ACM-14} & 2008 & $\blacksquare$ & $\square$ & $\square$ & $\square$ & $\blacksquare$ & $\square$ & $\square$\\
\cite{IEEE-163} & 2009 & $\square$ & $\blacksquare$ & $\square$ & $\square$ & $\blacksquare$ & $\square$ & $\square$\\
\cite{SB-21} & 2009 & $\blacksquare$ & $\square$ & $\square$ & $\square$ & $\blacksquare$ & $\square$ & $\square$\\
\cite{SB-70} & 2009 & $\blacksquare$ & $\square$ & $\square$ & $\square$ & $\square$ & $\square$ & $\blacksquare$\\
\cite{ACM-12} & 2010 & $\square$ & $\blacksquare$ & $\square$ & $\square$ & $\blacksquare$ & $\square$ & $\square$\\
\cite{SB-88} & 2010 & $\blacksquare$ & $\square$ & $\square$ & $\square$ & $\blacksquare$ & $\square$ & $\square$\\
\cite{IEEE-Penz2011CSEET} & 2011 & $\square$ & $\square$ & $\blacksquare$ & $\square$ & $\blacksquare$ & $\square$ & $\square$\\
\cite{IEEE-171} & 2011 & $\square$  & $\blacksquare$ & $\square$ & $\square$ & $\blacksquare$ & $\square$ & $\square$\\
\cite{SB-35} & 2011 & $\square$ & $\square$ & $\square$ & $\blacksquare$ & $\square$ & $\square$ & $\square$\\
\cite{SB-99} & 2011 & $\blacksquare$ & $\square$ & $\square$ & $\square$ & $\square$ & $\square$ & $\blacksquare$\\
\cite{ACM-11} & 2012 & $\square$ & $\blacksquare$ & $\square$ & $\square$ & $\blacksquare$ & $\square$ & $\square$\\
\cite{SB-07} & 2012 & $\square$ & $\square$ & $\square$ & $\square$ & $\blacksquare$ & $\square$ & $\square$\\
\cite{IEEE-Penz2013cseet} & 2013 & $\square$ & $\blacksquare$ & $\square$ & $\square$ & $\square$ & $\square$ & $\square$\\
\cite{SB-03} & 2013 & $\blacksquare$ & $\square$ & $\square$ & $\square$ & $\blacksquare$ & $\square$ & $\square$\\
\cite{SB-16} & 2013 & $\blacksquare$ & $\square$ & $\square$ & $\square$ & $\square$ & $\square$ & $\blacksquare$\\
\cite{SB-57} & 2013 & $\square$ & $\blacksquare$ & $\square$ & $\square$ & $\blacksquare$ & $\square$ & $\square$\\
\cite{IJSHE-13} & 2014 & $\blacksquare$ & $\square$ & $\square$ & $\square$ & $\blacksquare$ & $\square$ & $\square$\\
\cite{IEEE-147} & 2014 & $\square$ & $\blacksquare$ & $\square$ & $\square$ & $\blacksquare$ & $\square$ & $\square$\\
\cite{IEEE-165} & 2014 & $\square$ & $\blacksquare$ & $\square$ & $\square$ & $\blacksquare$ & $\square$ & $\square$\\
\cite{ACM-10} & 2014 & $\blacksquare$ & $\square$ & $\square$ & $\square$ & $\blacksquare$ & $\square$ & $\square$\\
\cite{SB-04} & 2014 & $\square$ & $\blacksquare$ & $\square$ & $\square$ & $\square$ & $\square$ & $\blacksquare$\\
\cite{SB-40} & 2014 & $\square$ & $\blacksquare$ & $\square$ & $\square$ & $\square$ & $\blacksquare$ & $\blacksquare$\\
\cite{SB-45} & 2014 & $\square$ & $\blacksquare$ & $\square$ & $\square$ & $\square$ & $\blacksquare$ & $\blacksquare$\\
\cite{SB-69} & 2014 & $\blacksquare$ & $\square$ & $\square$ & $\square$ & $\blacksquare$ & $\square$ & $\square$\\
\cite{IJSHE-6} & 2015 & $\blacksquare$ & $\square$ & $\square$ & $\square$ & $\blacksquare$ & $\square$ & $\square$\\
\cite{IJSHE-21} & 2015 & $\blacksquare$ & $\square$ & $\square$ & $\square$ & $\blacksquare$ & $\square$ & $\square$\\
\cite{Scopus-1} & 2015 & $\square$ & $\square$ & $\blacksquare$ & $\square$ & $\square$ & $\blacksquare$ & $\square$\\
\cite{IEEE-145} & 2015 & $\square$ & $\blacksquare$ & $\square$ & $\square$ & $\square$ & $\blacksquare$ & $\square$\\
\cite{SB-01} & 2015 & $\blacksquare$ & $\square$ & $\square$ & $\square$ & $\square$ & $\square$ & $\blacksquare$\\
\cite{SB-31} & 2015 & $\square$ & $\blacksquare$ & $\square$ & $\square$ & $\square$ & $\blacksquare$ & $\blacksquare$\\
\cite{IEEE-161} & 2016 & $\square$ & $\blacksquare$ & $\square$ & $\square$ & $\square$ & $\blacksquare$ & $\square$\\
\cite{IEEE-172} & 2016 & $\square$ & $\blacksquare$ & $\square$ & $\square$ & $\blacksquare$ & $\square$ & $\square$\\
\cite{SB-05} & 2016 & $\square$ & $\blacksquare$ & $\square$ & $\square$ & $\blacksquare$ & $\blacksquare$ & $\blacksquare$\\
\cite{SB-14} & 2016 & $\blacksquare$ & $\square$ & $\square$ & $\square$ & $\square$ & $\square$ & $\blacksquare$\\
\cite{SB-19} & 2016 & $\blacksquare$ & $\square$ & $\square$ & $\square$ & $\square$ & $\blacksquare$ & $\square$\\
\cite{SB-64} & 2016 & $\square$ & $\blacksquare$ & $\square$ & $\square$ & $\blacksquare$ & $\square$ & $\square$\\
\cite{SB-75} & 2016 & $\square$ & $\blacksquare$ & $\square$ & $\square$ & $\blacksquare$ & $\square$ & $\square$\\
\cite{SB-92} & 2016 & $\square$ & $\blacksquare$ & $\square$ & $\square$ & $\blacksquare$ & $\square$ & $\square$\\
\cite{WOS-5} & 2017 & $\blacksquare$ & $\square$ & $\square$ & $\square$ & $\square$ & $\blacksquare$ & $\square$\\
\cite{IEEE-153} & 2017 & $\square$ & $\blacksquare$ & $\square$ & $\square$ & $\blacksquare$ & $\square$ & $\square$\\
\cite{ACM-07} & 2017 & $\square$ & $\square$ & $\blacksquare$ & $\square$ & $\blacksquare$ & $\square$ & $\square$\\
\cite{SB-32} & 2017 & $\square$ & $\blacksquare$ & $\square$ & $\square$ & $\square$ & $\blacksquare$ & $\square$\\
\cite{IJSHE-11} & 2018 & $\blacksquare$ & $\square$ & $\square$ & $\square$ & $\blacksquare$ & $\square$ & $\square$\\
\cite{WoS-2} & 2018 & $\blacksquare$ & $\square$ & $\square$ & $\square$ & $\square$ & $\square$ & $\blacksquare$\\
\cite{IEEE-167} & 2018 & $\square$ & $\blacksquare$ & $\square$ & $\square$ & $\square$ & $\blacksquare$ & $\square$\\
\cite{ACM-06} & 2018 & $\square$ & $\square$ & $\blacksquare$ & $\square$ & $\square$ & $\blacksquare$ & $\square$\\
\cite{SB-77} & 2018 & $\blacksquare$ & $\square$ & $\square$ & $\square$ & $\square$ & $\blacksquare$ & $\square$\\
\cite{IJSHE-10} & 2019 & $\blacksquare$ & $\square$ & $\square$ & $\square$ & $\blacksquare$ & $\square$ & $\square$\\
\cite{CompSusEd-01} & 2019 & $\square$ & $\square$ & $\blacksquare$ & $\square$ & $\blacksquare$ & $\square$ & $\square$\\
\cite{CompSusEd-02} & 2019 & $\square$ & $\square$ & $\blacksquare$ & $\square$ & $\blacksquare$ & $\square$ & $\square$\\
\cite{CompSusEd-03} & 2019 & $\square$ & $\square$ & $\blacksquare$ & $\square$ & $\blacksquare$ & $\square$ & $\square$\\
\cite{IEEE-170} & 2019 & $\square$ & $\blacksquare$ & $\square$ & $\square$ & $\square$ & $\blacksquare$ & $\square$\\
\cite{IEEE-175} & 2019 & $\square$ & $\square$ & $\square$ & $\blacksquare$ & $\blacksquare$ & $\square$ & $\square$\\
\cite{SB-08} & 2019 & $\blacksquare$ & $\square$ & $\square$ & $\square$ & $\blacksquare$ & $\square$ & $\square$\\
\cite{SB-15} & 2019 & $\blacksquare$ & $\square$ & $\square$ & $\square$ & $\blacksquare$ & $\square$ & $\square$\\
\cite{SB-33} & 2019 & $\blacksquare$ & $\square$ & $\square$ & $\square$ & $\square$ & $\square$ & $\blacksquare$\\
\cite{SB-44} & 2019 & $\square$ & $\blacksquare$ & $\square$ & $\square$ & $\square$ & $\blacksquare$ & $\blacksquare$\\
\cite{SB-62} & 2019 & $\square$ & $\blacksquare$ & $\square$ & $\square$ & $\square$ & $\blacksquare$ & $\blacksquare$\\
\cite{SB-68} & 2019 & $\square$ & $\square$ & $\square$ & $\blacksquare$ & $\square$ & $\blacksquare$ & $\square$\\
\cite{IJSHE-8} & 2020 & $\blacksquare$ & $\square$ & $\square$ & $\square$ & $\blacksquare$ & $\square$ & $\square$\\
\cite{IJSHE-12} & 2020 & $\blacksquare$ & $\square$ & $\square$ & $\square$ & $\blacksquare$ & $\square$ & $\square$\\
\cite{IJSHE-26} & 2020 & $\blacksquare$ & $\square$ & $\square$ & $\square$ & $\blacksquare$ & $\square$ & $\square$\\
\cite{WoS-1} & 2020 & $\blacksquare$ & $\square$ & $\square$ & $\square$ & $\square$ & $\blacksquare$ & $\square$\\
\cite{ACM-01} & 2020 & $\square$ & $\blacksquare$ & $\square$ & $\square$ & $\square$ & $\blacksquare$ & $\blacksquare$\\
\cite{SB-29} & 2020 & $\blacksquare$ & $\square$ & $\square$ & $\square$ & $\square$ & $\square$ & $\blacksquare$\\
\cite{SB-46} & 2020 & $\square$ & $\blacksquare$ & $\square$ & $\square$ & $\square$ & $\blacksquare$ & $\square$\\
\cite{SB-53} & 2020 & $\blacksquare$ & $\square$ & $\square$ & $\square$ & $\blacksquare$ & $\square$ & $\square$\\
\cite{SB-54} & 2020 & $\blacksquare$ & $\square$ & $\square$ & $\square$ & $\blacksquare$ & $\square$ & $\square$\\
\cite{SB-56} & 2020 & $\blacksquare$ & $\square$ & $\square$ & $\square$ & $\square$ & $\blacksquare$ & $\square$\\
\cite{SB-71} & 2020 & $\blacksquare$ & $\square$ & $\square$ & $\square$ & $\square$ & $\square$ & $\blacksquare$\\
\cite{SB-82} & 2020 & $\blacksquare$ & $\square$ & $\square$ & $\square$ & $\square$ & $\square$ & $\blacksquare$\\
\cite{SB-55} & 2021 & $\blacksquare$ & $\square$ & $\square$ & $\square$ & $\square$ & $\square$ & $\blacksquare$\\
\cite{SB-79} & 2021 & $\square$ & $\square$ & $\square$ & $\blacksquare$ & $\square$ & $\blacksquare$ & $\square$\\
\cite{SB-89} & 2021 & $\blacksquare$ & $\square$ & $\square$ & $\square$ & $\blacksquare$ & $\square$ & $\square$\\
\cite{ACM-New-1} & 2021 & $\square$ & $\blacksquare$ & $\square$ & $\square$ & $\square$ & $\blacksquare$ & $\square$\\
\cite{ACM-New-2} & 2021 & $\square$ & $\blacksquare$ & $\square$ & $\square$ & $\blacksquare$ & $\square$ & $\square$\\
\cite{IEEE-New-1} & 2021 & $\square$ & $\square$ & $\blacksquare$ & $\square$ & $\square$ & $\blacksquare$ & $\square$\\
\cite{IEEE-New-2} & 2021 & $\square$ & $\blacksquare$ & $\square$ & $\square$ & $\blacksquare$ & $\square$ & $\square$\\
\cite{IEEE-New-3} & 2020 & $\square$ & $\blacksquare$ & $\square$ & $\square$ & $\blacksquare$ & $\square$ & $\square$\\
\cite{IEEE-New-4} & 2020 & $\square$ & $\blacksquare$ & $\square$ & $\square$ & $\blacksquare$ & $\square$ & $\square$\\
\cite{WoS-new-1} & 2020 & $\square$ & $\blacksquare$ & $\square$ & $\square$ & $\blacksquare$ & $\square$ & $\square$\\
\cite{IEEE-156} & 2010 & $\square$ & $\blacksquare$ & $\square$ & $\square$ & $\blacksquare$ & $\square$ & $\square$\\
\cite{SB-12} & 2010 & $\blacksquare$ & $\square$ & $\square$ & $\square$ & $\blacksquare$ & $\square$ & $\blacksquare$\\
\cite{IJSHE-7} & 2010 & $\blacksquare$ & $\square$ & $\square$ & $\blacksquare$ & $\square$ & $\blacksquare$ & $\square$\\
\cite{SB-87} & 2010 & $\square$ & $\square$ & $\blacksquare$ & $\square$ & $\blacksquare$ & $\blacksquare$ & $\square$\\
\cite{IEEE-164} & 2010 & $\square$ & $\blacksquare$ & $\square$ & $\square$ & $\blacksquare$ & $\blacksquare$ & $\blacksquare$\\
\end{longtable}


